\documentclass[structabstract]{aa}

\usepackage{graphicx}
\usepackage{txfonts}
\begin{document}
   \title{Modulated Gamma-ray emission from compact millisecond pulsar binary systems}


   \author{W. Bednarek
          }

   \institute{Department of Astrophysics, University of \L \'od\'z,
              ul. Pomorska 149/153, 90-236 \L \'od\'z, Poland\\
              \email{bednar@uni.lodz.pl}
             }

   \date{Received ; accepted }


\abstract
{{\bf A significant amount of the millisecond pulsars has been discovered within binary systems.  Tens of these millisecond pulsars emit $\gamma$-rays which are modulated with the pulsar period since this emission is produced in the inner pulsar magnetosphere. In several such binary systems the masses of the companion stars have been derived allowing to distinguish two classes of objects, called the Black Widow and the Redback binaries. Pulsars in these binary systems are expected to produce winds which, colliding with stellar winds, create conditions for acceleration of electrons. These electrons should interact with the anisotropic radiation from the companion stars producing $\gamma$-ray emission modulated with the orbital period of the binary system, similarly as observed in the massive TeV $\gamma$-ray binary systems.}}
{{\bf We consider the interaction of a  millisecond pulsar (MSP) wind with a very inhomogeneous stellar wind from the companion star within binary systems of the Black Widow and Redback types. Our aim is to determine the features of $\gamma$-ray emission produced in the collision region of the winds from a few example MSP binary systems.}}
{{\bf It is expected that the pulsar wind should mix efficiently with the inhomogeneous stellar wind. The mixed winds move outside the binary with a relatively low velocity.
Electrons accelerated in such mixed, turbulent winds can interact with the magnetic field and also strong radiation from the companion star producing not only synchrotron radiation but also $\gamma$-rays in the the Inverse Compton process fluxes of which are expected to be modulated on the periods of the binary systems. Applying numerical methods, we calculated the GeV-TeV gamma-ray spectra and the light curves expected from some millisecond pulsar binary systems.}}
{{\bf Gamma-ray emission, produced within the binary systems, is confronted with the sensitivities of the present and future gamma-ray telescopes. It is concluded that energetic millisecond pulsar binary systems create a new class of TeV $\gamma$-ray sources which could be detectable by the future Cherenkov arrays (e.g. CTA) and possibly also by the extensive campains with the present arrays (HESS, MAGIC, VERITAS). However, $\gamma$-ray emission from the millisecond pulsar binary systems is predicted to have different features than those observed in the case of massive TeV gamma-ray binaries such as LS I 303 61 or LS 5039. The maximum in the TeV $\gamma$-ray orbital light curve should appear when the MSP is behind the companion star. This is in contrary to the observations of the orbital light curves from the massive TeV $\gamma$-ray binaries (LS I 303 61 or LS 5039). Moreover, the GeV and orbital TeV $\gamma$-ray light curves should be positively correlated in contrary to the case of massive TeV $\gamma$-ray binaries.}}
{{\bf We conclude that TeV $\gamma$-ray emission, modulated on the orbital period of MSP binary systems, should be detected by the future CTA. Moreover, some MSP binary systems of the Redback type might also show GeV $\gamma$-ray emission modulated on the binary periods on the level detectable by Fermi-LAT.}}

\maketitle
%

%
%
\section{Introduction}

Most of the millisecond pulsars are members of binary systems (Manchester et al.~2005). It is supposed that this feature is related to their formation mechanism (Alpar et al.~1982; Bhattacharya \& van den Heuvel~1991). Several such binaries have surprisingly low mass companion stars. Their masses are of the order of a few percent of the solar mass in the case of Black Widow binaries and a few tens percent of solar mass in the case of Redback binaries (Roberts~2012). Clearly lower masses of the companion stars in Black Widow systems in respect to Redback systems are probably due to the more effective evaporation of the companion stars caused by the energy released by the MSPs in these binaries (Ruderman et al.~1989). The energy realised by the millisecond pulsar also heats the surface of the companion star to temperature significantly above that expected from the nuclear burning in Black Widow systems. Therefore, optical emission from the companions in Black Widow systems is often modulated on the periods of the binary systems (e.g. Breton et al.~2013), in contrast to the emission from the companion stars in the Redback systems.

As observed in the famous binary system containing the pulsar PSR B1259$-$63, it is expected that electrons can also be accelerated within the MSP binaries at the shock which appears to be a result of the interaction of winds from the millisecond pulsar and the companion star. Electrons can reach  energies allowing them to produce of synchrotron X-rays and Inverse Compton $\gamma$-rays in collisions with the radiation from the companion stars. One of the best studied MSP binary system of the Black Widow type is PSR B1957+20 (Fruchter et al.~1988). The X-ray emission from this binary system is modulated on the orbital period (Huang et al.~2012), suggesting its origin in the wind collision shock as considered by Harding \& Gaisser (1990) and Arons \& Tavani~(1993). In fact, the presence of the shock within the binary system has been expected from the discovery of binary system PSR B1957+20, since the pulsed radio emission shows the eclipse lasting for about 10$\%$ of the orbital phase (Fruchter et al.~1988). The binary system PSR B1957+20 was claimed in the past to be a GeV-TeV $\gamma$-rays source (Brink et al. 1990). But, this report was not confirmed by EGRET observations (Buccheri et al. 1996).
Only recently, {\it Fermi} discovered pulsed $\gamma$-ray emission from PSR B1957+20 (Guillemot et al.~2012). It is clear at present that MSPs create one of the major class of pulsed $\gamma$-ray sources
in the Galaxy. There are also evidences of additional higher energy component in the GeV emission from the Black Widow binary system PSR B1957+20 (Wu et al.~2012). This component is modulated on the period of the binary system reaching the maximum when the MSP is behind the companion star. This additional emission component, modulated on the orbital period of the binary system, is interpreted as produced by relativistic electrons moving in the pulsar wind (Wu et al.~2012). They up-scatter radiation field from the companion star. 

Similar general features of the pulsed GeV $\gamma$-ray emission from the MSPs and classical pulsars (efficiencies and shape of the spectra) suggest that
acceleration processes in the inner magnetospheres of these two classes of pulsars occurs similarly (Abdo et al.~2009a, Abdo et al.~2013).
Therefore, it is likely that binary systems containing MSPs also show modulated high energy emission, as recently discovered in the case of TeV $\gamma$-ray binaries such as those  containing PSR B1259$-$63 (Abdo et al.~2011, Aharonian et al.~2005a) and maybe also LS 5039 (Abdo et al.~2009b; Aharonian et al.~2005b) and LSI 61 303 (Abdo et al.~2009c; Albert et al.~2006a), provided that these two binary systems contain energetic pulsars as suggested by e.g. Dubus~(2006).

In this paper we consider a specific model for the interaction of winds from the MSP and the companion star. In contrast to previous models (Harding \& Gaisser 1990, Arons \& Tavani~1993), we assume that the wind from the companion star is highly inhomogeneous. Then, the pulsar and stellar winds should mix very efficiently at the location where the pressure of both winds becomes comparable.  The mixed winds expand from the binary system with a relatively low velocity due to the barion loading of the relativistic pulsar wind with the matter from the relatively slow stellar wind. Relativistic electrons, captured in such mixed winds, can interact efficiently with the radiation from the nearby companion star producing TeV $\gamma$-rays, which are supposed to be modulated on the orbital period of the binary system. We calculate the $\gamma$-ray and synchrotron spectra, produced by these electrons, and confront them with the available X-ray observations from these binary systems and also with the sensitivities of the Fermi-LAT telescope and the  present and future Cherenkov telescopes.

\section{Interaction of winds in millisecond pulsar binary systems}

We consider the standard picture for the compact binary systems containing energetic millisecond pulsars
of the Black Widow and Redback type following previous works by Harding \& Gaisser~(1990) and Arons \& Tavani~(1993). 
In this scenario  a millisecond pulsar produces a relativistic pulsar wind which interacts with the stellar wind of the low mass companion (see Fig.~1 on the left). As a result of this interaction, a shock structure appears within the binary system which separates the pulsar and stellar winds. We modify this standard scenario by assuming that the wind from the companion star is highly inhomogeneous. In such case, the interaction of winds produce very turbulent mixed pulsar-stellar wind which moves together with the velocity determined by the momentum conservation. The velocity of the mixed pulsar-stellar wind can be estimated by assuming that the relativistic pulsar wind is loaded with the matter from the stellar wind,
\begin{eqnarray}
v_{\rm mix} = [{{2L_{\rm pul}\Delta\Omega_{\rm pul}}\over{{\dot M}_\star\Delta\Omega_{\star}}}]^{1/2}\approx 4\times 10^9({{\xi_{-1}L_{35}}\over{M_{-11}}})^{1/2}~~~{\rm {{cm}\over{s}}}
\label{eq1}
\end{eqnarray}
\noindent
where $L_{\rm pul} = 10^{35}L_{35}$ erg s$^{-1}$ is the power of the pulsar wind, and $\dot{M} = 10^{-11}M_{-11}$ M$_\odot$ yr$^{-1}$ is the mass loss rate of the companion star. 
The power of the pulsar wind is assumed to be equal to the rotational energy loss rate of the pulsar which is estimated for known values of the rotational period, $P$, and period derivative of the pulsar, $\dot{P}$, based on the standard formula which assume the rotating dipole model for the pulsar magnetosphere: $L_{\rm pul} = 3.95\times 10^{31} (\dot{\rm P}/10^{-15})({\rm P/1 s})^{-3}~{\rm erg~ s}^{-1}$.
$\Delta\Omega_{\rm pul}$ and  $\Delta\Omega_{\star}$ describe a part of the solid angles which are over-taken by the collision region as observed from the pulsar site and from the companion star site, respectively. 
These solid angles can be estimated as $\Delta\Omega_{\rm pul} = (1 - \cos\phi_{\rm pul})/2$ and $\Delta\Omega_{\star} = (1 - \cos\phi_\star)/2$. We introduced the parameter $\chi = 0.1\chi_{-1} = \Delta\Omega_{\rm pul}/\Delta\Omega_{\star}$.
$\phi_{\rm pul}$ is the half opening angle calculated from $\phi = 2.1(1 - \eta^{2/5}/4)\eta^{1/3}$ rad  (see Eichler \& Usov~1993). $\eta$ is the minimum value from ${\eta_{\rm 0}, \eta_{\rm 0}^{-1}}$. $\eta_{\rm 0}$ defined by the parameters of
the pulsar and stellar winds $\eta_{\rm 0} = (L_{\rm pul}/c)/\dot{M}_\star v_\star\approx 5L_{35}/(M_{-10}v_8)$, and the velocity of the stellar wind is $v_\star = 10^8v_8$ cm s$^{-1}$ (Girard \& Willson~1987).  $\phi_\star$ is the half opening angle of the collision region seen from the location of the companion star.
The closest approach of the shock from the companion star is at $\rho_\star = D_{\rm b}/(1 + \sqrt{\eta})$, where $D_{\rm b}$ is the separation of the stars. The distance of the shock for other angles, $\phi_\star$, is approximated by $\rho(\phi_\star)\approx \rho_\star\times \phi_\star/\sin (\phi_\star)$. We estimate the angle $\phi_\star$ by solving the approximate equation: 
$D_{\rm b} \cos \phi_\star \approx \rho_\star (\phi_\star/\sin \phi_\star)$.
For typical parameters of the millisecond pulsar binary systems
($L = 10^{35}$ erg s$^{-1}$, $M = 10^{-11}$ M$_\odot$ yr$^{-1}$ and $v_\star = 700$ km s$^{-1}$, expected for PSR B1957+20), $\eta$ is estimated on $\sim$7, the angle $\phi_{\rm pul}$ on $\sim$0.45 rad and the angle
$\phi_\star$ on 1.2 rad. Then, the value of the parameter $\xi$ is equal to $\sim 0.14$. For these parameters, typical velocity of the mixed pulsar/stellar wind is $\sim 4.6\times 10^9$ cm s$^{-1}$. 

\begin{figure*}[t]
\vskip 8.5truecm
\includegraphics{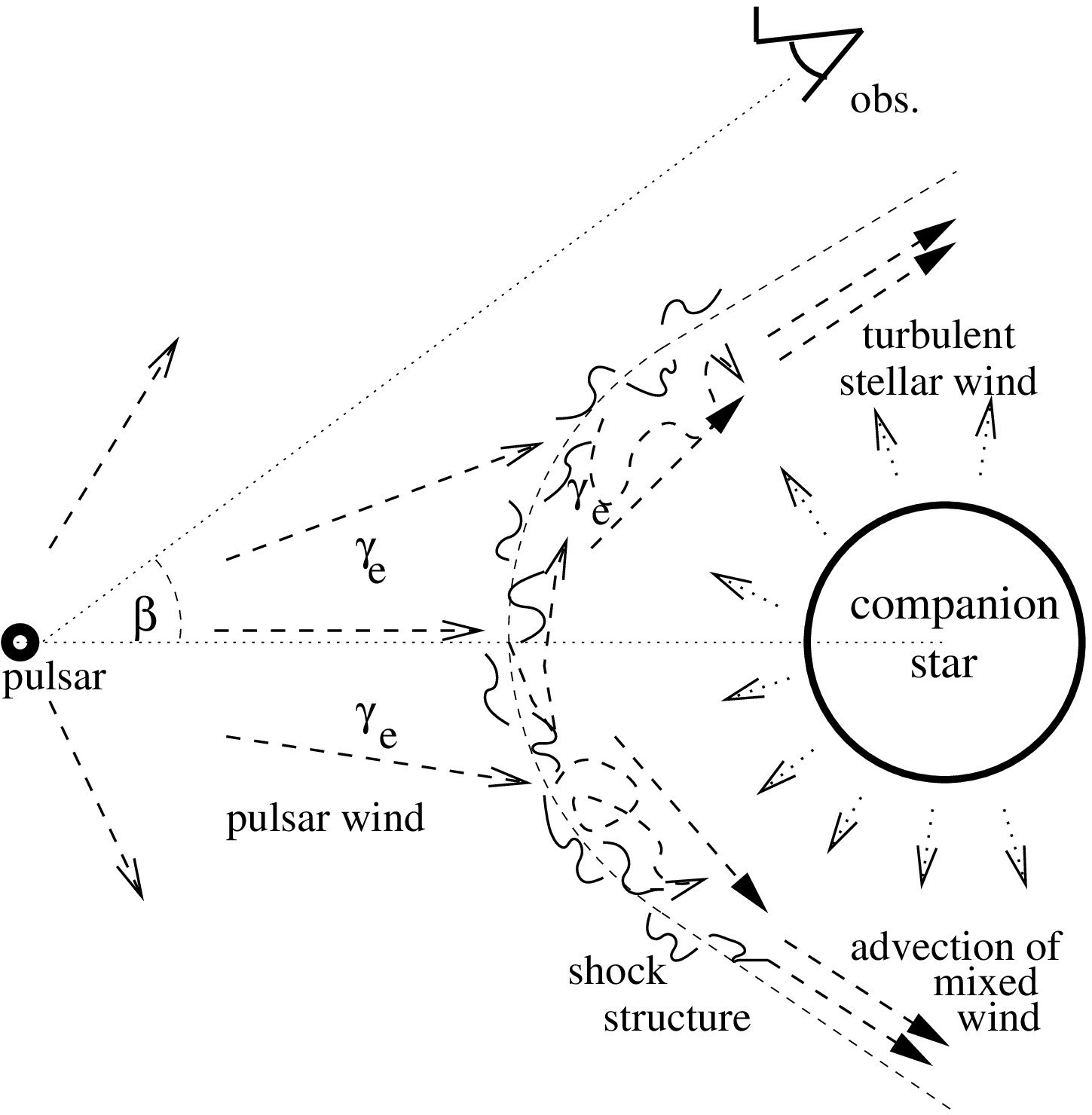}
\includegraphics{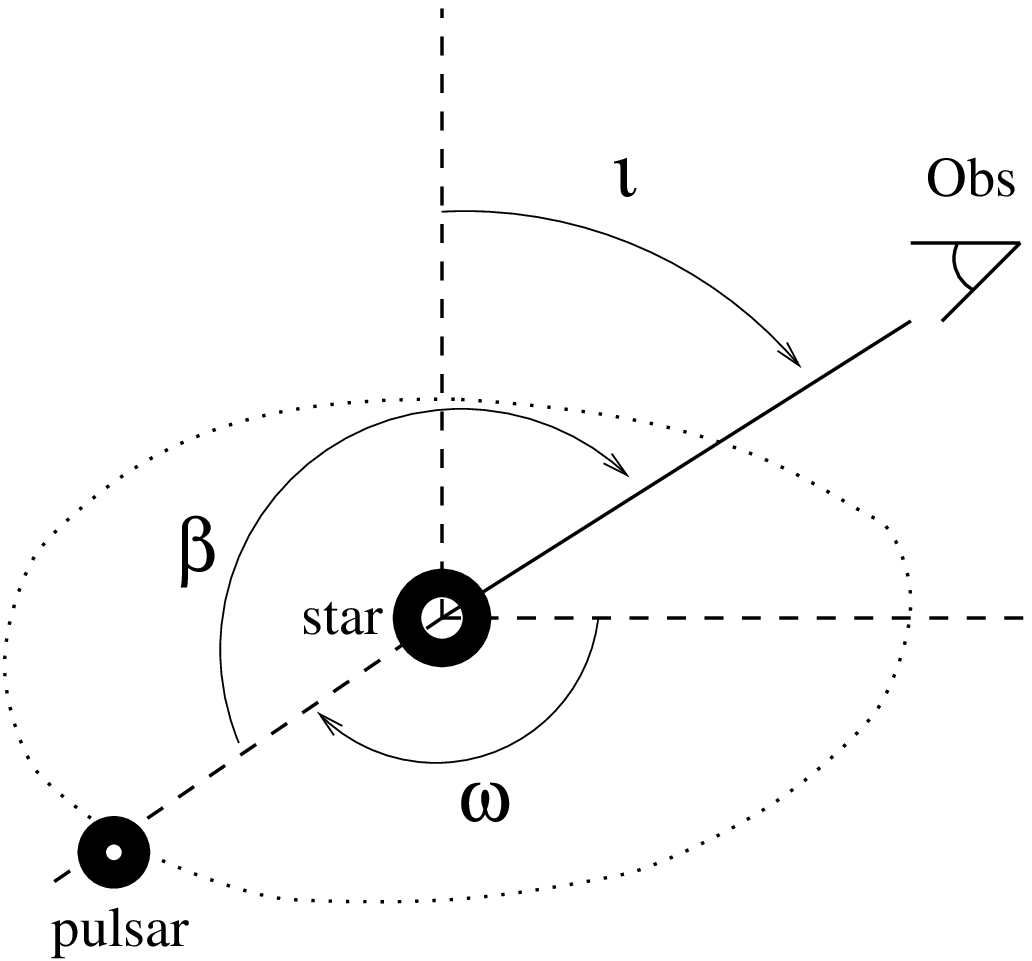}
\caption{Schematic representation of the compact binary system containing energetic millisecond pulsar (on the left) and the picture on which the basic angles defining the binary system are shown (on the right). On the left: The pulsar creates a strong wind which contains relativistic electrons ($\gamma_{\rm e}$).
The companion star creates turbulent stellar wind mainly as a result of irradiation from the pulsar. The winds collide creating turbulent, non-stationary shock structure. Both winds mix effectively in this region. Relativistic electrons (from the pulsar wind or additionally re-accelerated in the turbulent pulsar-stellar winds) are isotropized by the random magnetic field of the mixed winds.
They are advected along the shock structure. During this process, electrons comptonize stellar radiation producing $\gamma$-rays. Due to the anisotropy of the stellar radiation field,  $\gamma$-rays are produced preferentially in general direction towards the companion star. 
As a result, the $\gamma$-ray signal, produced towards the observer (obs.)  located at the angle $\beta$, should be modulated with the phase of the binary system.
On the right: The binary system is inclined at the angle $i$ towards the observer's (Obs) line of sight. The orbital phase, $\omega$, is equal to zero when the pulsar is in front of the companion star. The angle between the observer and the pulsar (and also injection place of relativistic electrons) is marked by $\beta$.}
\label{fig1}
\end{figure*}

The above estimated relatively low velocity of the mixed winds (in respect to the velocity of the pulsar wind) guarantee that relativistic electrons stay relatively close to the stellar surface for long enough time that their interactions with soft radiation from the stellar surface is efficient. These electrons are isotropized in the reference frame of the mixed wind. They lose energy mainly on the synchrotron process and the IC scattering of stellar radiation. 

In Table.~1, we report the basic parameters of a few MSP binary systems. Up till now, several MSP binary systems have been identified to belong to the Black Widow type and a few such systems has  been classified as the Redback type (Roberts~2012). For the illustration purposes, we have selected two MSP binaries of the Black Widow type, PSR B1957+20 (Fruchter et al.~1988) and PSR J1810+1744 (Hessels et al.~2011), and two of the Redback type, PSR J1023+0038 (Archibald et al.~2009) and PSR J1816+4510  (Kaplan et al.~2012). 
The basic parameters of these MSP binaries are quite well known. The distances to these binaries are estimated based on the dispersion measure using the Cordes \& Lazio (2002) model (see Gentile et al.~2013) and from parallax measurements for PSR 1023+0038 (Deller et al.~2012). The parameters of the companion stars (stellar radius, and surface temperature) and the parameters of the binary systems (semimajor axis, inclination angle) are taken from Manchester et al.~(2005), Fruchter et al.~(1990), Reynolds et al.~(2007), Archibald et al.~(2009), Hessels et al.~(2011), Breton et al.(2013), and Kaplan et al.~(2012). Note that the inclination angles of these binary systems are not very well defined since their estimates depend on the unknown masses of the neutron stars in these binary systems.
These four binaries are expected to provide good conditions for the $\gamma$-ray production by electrons accelerated at the collision region of the pulsar and stellar winds. These binary systems are compact, contain MSPs with large rotational energy loss rate (estimated for the known rotational periods and period derivatives, see Manchester et al.~2005, Archibald et al.~2009  and Kaplan et al.~2012), the collision regions of their winds are close to the stellar surfaces and their measured surface temperatures are relatively high. The distances of the collision regions from the companion stars, $R_{\rm sh}$ (reported in Table~1), are estimated by the value of the closest approach of the collision region to the companion star, $\rho_\star$ (defined above).

\begin{table*}
  \caption{Basic parameters of millisecond pulsars and their binary systems:
radius of the companion star ($R_\star$) and its surface temperature ($T_\star$), semimajor axis of the orbit (a), distance of the shock from stellar surface ($R_{\rm sh}$), inclination of the binary system ($i$), the pulsar rotational period ($P_{\rm pul}$), the pulsar energy loss rate ($L_{\rm pul}$), distance to the binary system (D), and the X-ray luminosities from the binary systems are taken from: [1] Huang et al.~(2012), [2] Bogdanov et al.~(2011), [3] Kaplan et al.~(2012), and [4] Gentile et al.~(2013).}
  \begin{tabular}{llllllllllllll}
\hline 
\hline 
Name   &  $R_\star$ (cm) &  $T_\star$ (K) &  $a$ (cm)  &  $i$  &  $R_{\rm sh}$ ($R_\star$) & $P_{\rm pul}$ (ms)  & $L_{\rm pul}$ (erg/s) & D (kpc)  &  $F_{\rm x}$ (erg/cm$^2$/s) \\
\hline
PSR B1957+20 & $10^{10}$  &   $8\times 10^3$  &  $1.7\times 10^{11}$& 65$^\circ\pm 2^\circ$  & ~~~~5  & 1.61  & $1.6\times 10^{35}$  &  2.5  &  ~~~$6\times 10^{-14}$ [1] \\
PSR J1023+0038    &  $3\times 10^{10}$  &  $6.65\times 10^3$  &  $1.7\times 10^{11}$ & 34$^\circ - 53^\circ$  &  ~~~~2  &  1.69   &$1.2\times 10^{35}$  &  1.3  & ~~~$4\times 10^{-13}$ [2] \\
PSR J1816+4510    &  $8.4\times 10^{9}$  &  $2\times 10^4$  &  $1.2\times 10^{11}$ & ? & ~~~~4 & 3.19    &  $10^{35}$  &   2.4  & $< 3\times 10^{-14}$ [3]\\
PSR J1810+1744    &  $1.4\times 10^{10}$   &   $8\times 10^3$  &  $9.3\times 10^{10}$ & $48^\circ\pm  7^\circ$ &  ~~~~3  & 1.66   & $4\times 10^{34}$   &  2.  &  ~~~$2\times 10^{-14}$ [4] \\
\hline
\hline 
\end{tabular}
  \label{tab1}
\end{table*}

%
\section{High energy processes within binary systems}

We consider the role of the high energy processes which are expected to be important in the  discussed above general scenario.
We are interested in processes which could be able to explain X-ray emission produced within the MSP binary systems of the PSR B1957+20 type. This emission can not be produced by relativistic electrons which escape from the inner pulsar magnetosphere and move rectilinear in the pulsar wind.
On the other hand, such synchrotron X-ray emission can be produced within the binary system when the relativistic electrons move in the region of turbulent, mixed MSP-stellar wind discussed above. In fact, the phase dependent X-ray emission is observed up to at least $\varepsilon_{\rm syn}\sim$8 keV (Huang et al.~2012). We assume that this emission originates in the synchrotron process. In such case, the magnetic field in the emission region and the Lorentz factor of electrons should fulfil the following condition,
\begin{eqnarray}
\varepsilon_{\rm syn} = m_{\rm e}c^2(B/B_{\rm cr})\gamma_{\rm e}^2,
\label{eq2}
\end{eqnarray}
\noindent
where $m_{\rm e}c^2$ is the electron rest energy, $B$ is the magnetic field strength in the emission region, $B_{\rm cr} = 4.4\times 10^{13}$ G is the critical magnetic field strength, and $\gamma_{\rm e}$ is the Lorentz factor of electrons. This synchrotron emission is likely to originate at the transition region of colliding winds.
The magnetic field strength in this region can be estimated by a simple extrapolation from the pulsar surface, assuming the dipole structure of the magnetic field below the light cylinder and the toroidal structure in the pulsar wind region. Then, 
\begin{eqnarray}
B_{\rm sh} = 3\sigma^{1/2} B_{\rm pul}\left({{R_{\rm pul}}\over{R_{\rm LC}}}\right)^3{{R_{\rm LC}}\over{\rho_{\rm o}}}\approx {{0.3\sigma_{-4}^{1/2} B_8}\over{P_2^2\rho_{11}}}~~~{\rm G}, 
\label{eq3}
\end{eqnarray}
\noindent
where $B_{\rm pul} = 10^8B_8$ G and $P_{\rm pul} = 2P_2$ ms are the surface magnetic field strength and the period of the pulsar, $R_{\rm LC} = cP_{\rm pul}/2\pi$ is the light cylinder radius of the pulsar, $\sigma = 10^{-4}\sigma_{-4}$ is the magnetization parameter of the pulsar wind, and $\rho_{o} = a - R_{\rm sh}R_\star$ is the distance of the wind collision region from the MSP and $a$ is the semimajor axis of the binary system. The surface magnetic field of the pulsar is estimated based on the known values of the pulsar period and period derivative and 
the standard formula which assumes the rotating dipole model for the inner pulsar magnetosphere $B_{\rm pul} = 3.2\times 10^{19} ({\rm P {\dot P}})^{1/2}~{\rm G}$.
Note that the magnetic field is expected to reconnect very efficiently in this highly turbulent region of the mixed winds. Therefore, the magnetic field strengths, described by very low values of the magnetization parameter, $\sigma << 10^{-2}$, should not be surprising. For the above value of the magnetic field at the transition region (Eq.~\ref{eq3}), the Lorentz factor of electrons, have to be at least,
\begin{eqnarray}
\gamma_{\rm e}\sim 1.5\times 10^6 P_2\rho_{11}^{1/2}/(\sigma_{-4}^{1/4}B_8^{1/2}),
\label{eq4}
\end{eqnarray}
\noindent
in order to produce the observable X-ray emission in synchrotron process (Eq.~2).
We conclude that the TeV electrons have to be present within the turbulent region of the colliding winds of the MSP binary system PSR B1957+20, in order to explain the X-ray emission modulated with the period of the binary system.

Let us determine the conditions in which electrons can reach TeV energies.
The maximum energies of electrons accelerated at the transition region are determined by the 
acceleration time scale and the time scales for the energy losses or the escape from the acceleration region.
The acceleration time scale can be estimated from,
\begin{eqnarray}
\tau_{\rm acc} = R_{\rm L}/(c\chi )\approx 1 E/(\chi_{-1}B)~~~{\rm s}, 
\label{eq5}
\end{eqnarray}
\noindent
where the electron energy is in TeV and $\chi = 0.1\chi_{-1}$ is the acceleration parameter. We assume that the acceleration parameter can be related to the velocity of the mixed winds in the following way:
$\chi\sim (v_{\rm mix}/c)^2 = 0.1v_{10}^2$, where the mixed wind velocity is $v_{\rm mix} = 10^{10}v_{10}$ cm s$^{-1}$.

The maximum energies of electrons might be constrained by their escape from the acceleration region with the velocity of the mixed pulsar/stellar wind. The advection time scale of the mixed wind is,
\begin{eqnarray}
\tau_{\rm adv} = \pi R_{\rm sh}R_\star/v_{\rm mix}\approx 10R_{11}/v_{10}~~~{\rm s},
\label{eq6}
\end{eqnarray}
\noindent
where $R\approx \pi R_{\rm sh}R_\star = 10^{11}R_{11}$ cm is the characteristic distance scale for the  propagation of the turbulent wind around the companion star. By comparing this time scale with the acceleration time scale, we get the limit on the electron energies,
\begin{eqnarray}
E_{\rm adv}^{\rm max}\approx 3R_{11}\sigma_{-4}^{1/2}B_8v_{10}/(P_2^2\rho_{11})~~~{\rm TeV}. 
\label{eq7}
\end{eqnarray}
\noindent
We conclude that electrons can be accelerated to TeV energies in the turbulent collision region of the mixed pulsar and stellar winds. 

Let us consider energy losses of accelerated electrons. To fully understand their importance, we compare the electron acceleration time scale with the synchrotron or the Inverse Compton (IC) 
energy loss time scales. The synchrotron energy loss time scale can be calculated from,
\begin{eqnarray}
\tau_{\rm syn} = E_{\rm e}/\dot{E}_{\rm syn}\approx 370/(B^2E)~~~{\rm s}.
\label{eq8}
\end{eqnarray}
\noindent
The synchrotron energy losses dominate over the IC energy losses in the Thomson regime
(T regime) provided that the magnetic field in the acceleration region is stronger than $B\approx 40T_4^2/R_{\rm sh}$ G (see e.g. Bednarek~1997), where $R_{\rm sh}$ is the distance of the electron acceleration from the companion star in units of the stellar radius. 
For the parameters of considered binary systems (see Table~1), the IC energy losses in the Thomson regime usually dominate over the synchrotron energy losses. However, the IC energy losses in the Klein-Nishina (KN) regime 
declines and the synchrotron losses can start to dominate at large energies.
In such case, the maximum energy of accelerated electrons is determined by the balance between the acceleration time scale and the synchrotron time scale. They are estimated on, 
\begin{eqnarray}
E_{\rm syn}^{\rm max}\approx 19(\chi_{-1}/B)^{1/2}~~~{\rm TeV}. 
\label{eq9}
\end{eqnarray}
\noindent
Then, electrons can reach energies as large as (see Eq.~3),
\begin{eqnarray}
E_{\rm syn}^{\rm max}\approx 35 P_2v_{10}\rho_{11}^{1/2}/(\sigma_{-4}^{1/4}B_8^{1/2})~~~{\rm TeV}.
\label{eq10}
\end{eqnarray}
\noindent
In the situation considered above, electrons with maximum energies lose energy mainly on the synchrotron process but lower energy electrons can lose most of their energy still on the IC process. Let us estimate the electron energy at which synchrotron losses become comparable to the IC losses in the KN regime. For the IC energy loss time scale in the KN regime, we apply the approximate formula by assuming that the IC energy losses in KN regime are comparable to the IC energy losses on the border between T and KN regimes (see e.g. Eq.~7 in Bednarek~2011). Then, the IC losses in KN regime can be estimated from, 
\begin{eqnarray}
\tau_{\rm IC}^{\rm KN} = {{3m_{\rm e}^2c^4E}\over{4c\sigma_{\rm T}U_{\rm rad}E_{\rm T/KN}^2}}\approx {{16.7ER_{\rm sh}^2}\over{T_4^2}}~~~{\rm s},
\label{eq11}
\end{eqnarray}
\noindent
where $\sigma_{\rm T}$ is the Thomson cross section, $U_{\rm rad}$ is the density of stellar radiation at the acceleration region, $T = 10^4T_4$ K is the surface temperature of the companion star, and $E_{\rm T/KN} = (m_{\rm e}c^2)^2/3k_{\rm B}T\approx 0.1/T_4$ TeV is the electron energy corresponding to transition between the T and KN regimes. By comparing $\tau_{\rm IC}^{\rm KN}$ with 
$\tau_{\rm syn}$, we can estimate the critical energy below which electrons lose energy mainly  on the IC process,
\begin{eqnarray}
E_{\rm syn}^{\rm IC/KN}\approx 4.7{{T_4}\over{R_{\rm sh}B_{\rm sh}}}\approx 
16 {{T_4P_2^2\rho_{11}}\over{\sigma_{-4}^{1/2}B_8R_{\rm sh}}}~~~{\rm TeV}.
\label{eq12}
\end{eqnarray}
\noindent
This critical electron energy is usually not far from the maximum electron energy determined from the comparison of the acceleration time scale and the synchrotron energy loss time scale.
Therefore, we conclude that a relatively small part of electron's energy can go to the synchrotron emission in respect to the IC emission, provided  that electrons have the differential power law spectrum and spectral index close to -2.
The IC $\gamma$-ray emission should in such case extend up to TeV energies. The presence of electrons with TeV energies can also explain observations of the non-thermal synchrotron emission from the binary system PSR B1957+20 mentioned above.

Note that the Larmor radii of the TeV electrons are lower than the characteristic dimension of the binary system,
$R_{\rm L} < a$, for the magnetic field strength above $B\sim 0.03E/a_{11}$ G, where $a = 10^{11}a_{11}$. This condition is fulfilled in our scenario (see Eq.~3), which means that electrons are trapped in the turbulent region. They have to be advected with the mixed winds.

We conclude that the most important process, which determines the maximum energies of electrons accelerated in the turbulent region, is the advection from the acceleration region (Eq.~7). However,  electrons can lose significant amount of their energy on radiation processes such as the IC scattering of soft photons from the companion star and the synchrotron radiation in the magnetic field in the colliding wind region.

\section{Gamma-rays from comptonization of stellar radiation}

A few MSP binary systems are selected for their well known parameters of the pulsars and binary systems (shown in Table~1). We assume that electrons are accelerated in the turbulent collision region of the pulsar and stellar winds to TeV energies as discussed in the section above. These electrons are isotropized in the reference frame of the mixed pulsar-stellar wind. They are advected from the acceleration region (assumed the apex of the collision region) with the velocity of the mixed winds. Since electrons are isotropized, they can efficiently interact with the stellar radiation. The injection place of electrons is located outside the isotropic source of stellar radiation. Therefore, the IC scattering process has to occur anisotropically because the electron-photon collision rate depends on the angle between the direction of propagation of the electron and the soft photon. The production rate (and spectra) of $\gamma$-rays should depend on the observation angle of the binary system in respect to the direction defined by the stars. In order to calculate the $\gamma$-ray spectra expected in terms of our model, we apply the Monte Carlo code which follows the Inverse Compton $e^\pm$ pair cascade processes. This code has been developed for the anisotropic radiation processes within the massive binary systems (Bednarek~1997, 2000, 2006). In the modified version of this code, we take into account not only radiation processes characteristic for the IC $e^\pm$ pair cascade within the binary system but also energy losses of electrons on the synchrotron process.
Therefore, we calculate the $\gamma$-ray spectra from the IC process and the synchrotron spectra produced by these same electrons. In fact, in the case of MSP binary systems, in which radiation field is much weaker than within
massive binary systems, some production of secondary $e^\pm$ pairs, due to the absorption of IC $\gamma$-rays, can only happen  for $\gamma$-rays which appear close to the stellar surface. We note that the radiation effects, connected with the production of secondary $e^\pm$ pairs, are negligible in the case of MSP binary systems.

In the example calculations, we assume that electrons are injected with the differential power law spectrum (spectral index -2), extending to the maximum energies allowed by their escape from the binary system (Eq.~7). The $\gamma$-ray spectra, escaping from the binary system, are obtained as a function of the cosine angle $\beta$ ($\cos\beta$), where $\beta$ is the angle between the direction defined by the stars and the direction towards the observer. The angle $\beta$ 
depends on the inclination angle of the binary system $i$ and the phase, $\omega$, of the injection place of relativistic electrons according to $\cos\beta = \cos(90^\circ - \iota) \cdot \cos\omega$. These angles are defined in Fig.~1 (on the right).

The synchrotron spectra are also calculated. In contrast to the IC spectra, the synchrotron spectra are isotropic, since the magnetic field and electron distribution in the mixed wind reference frame are assumed to be isotropic. The velocity of the mixed wind is relatively low. So then, relativistic beaming effects can be safely neglected. The only modulation effect on the X-ray emission is due to its absorption in the inhomogeneous wind of the companion star as observed in the case of PSR B1957+20.

\begin{figure*}
\vskip 12.5truecm
\includegraphics{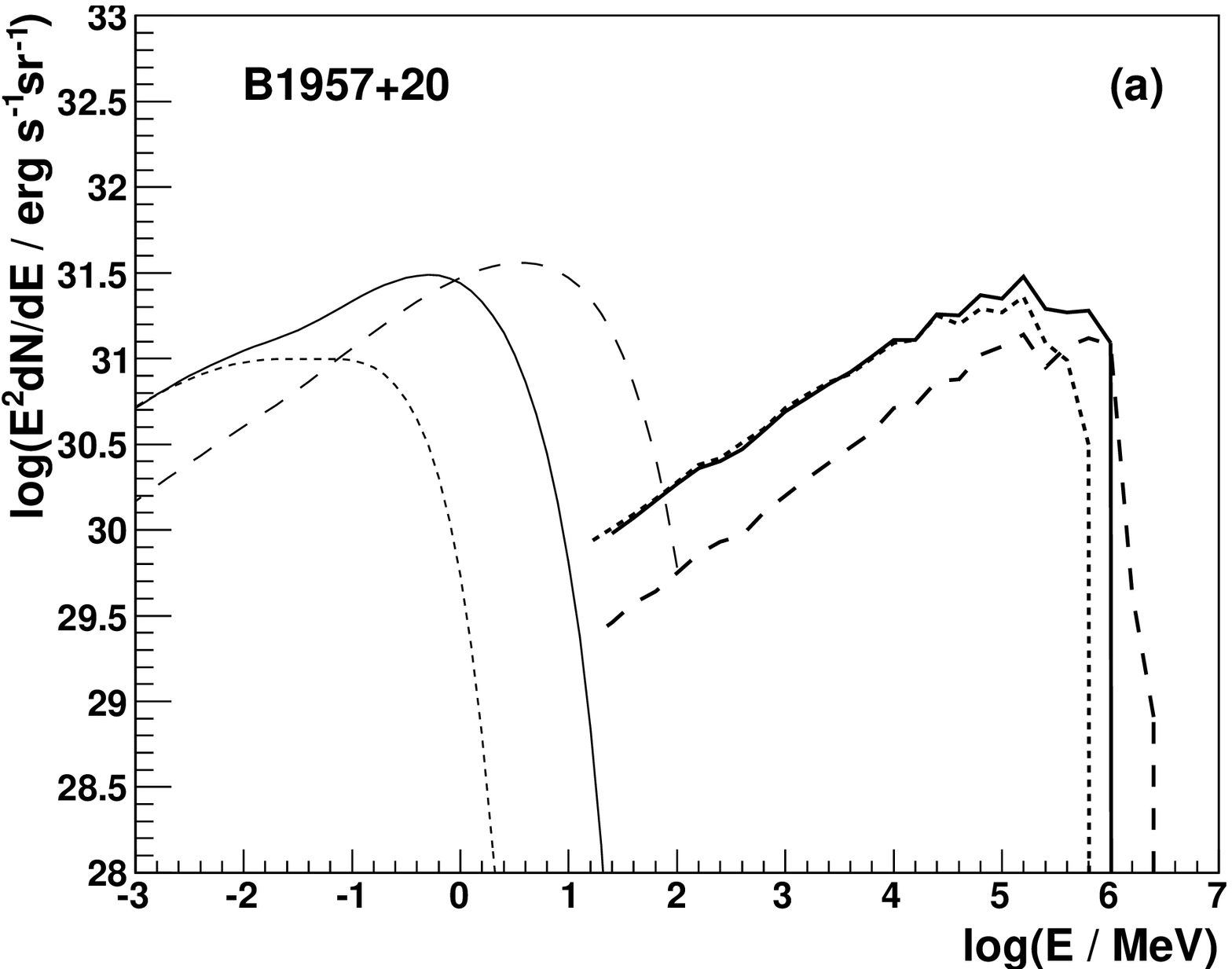}
\includegraphics{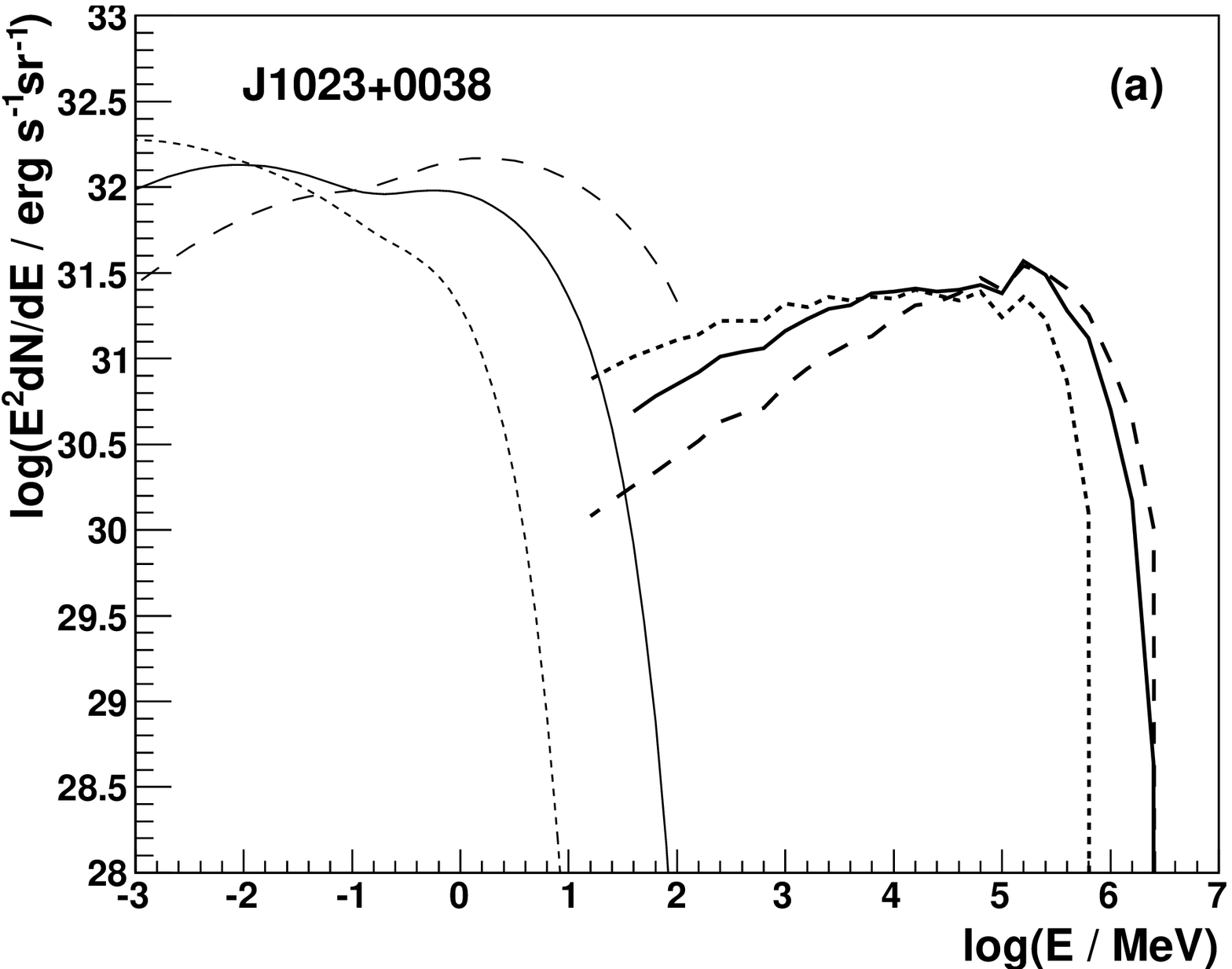}
\includegraphics{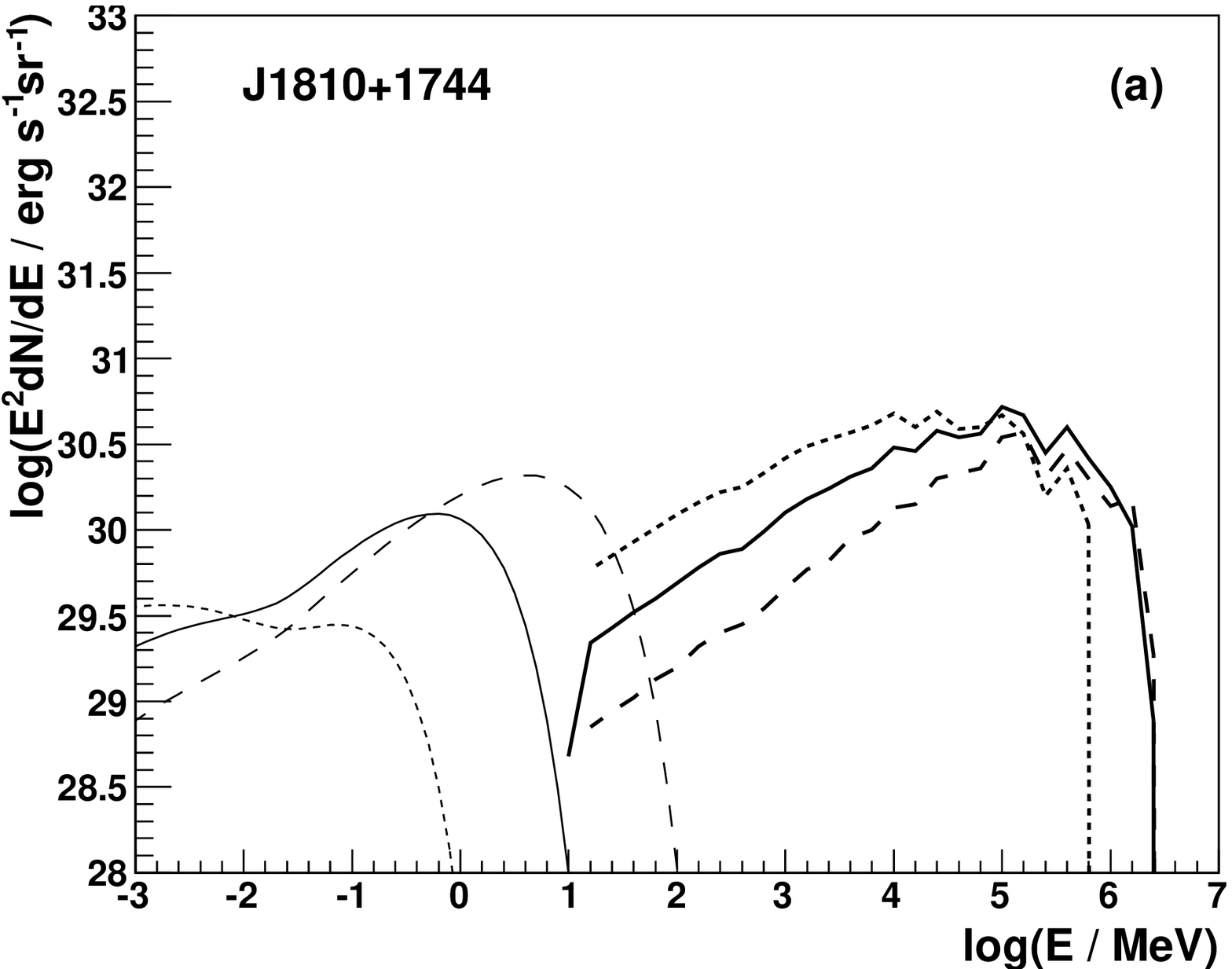}
\includegraphics{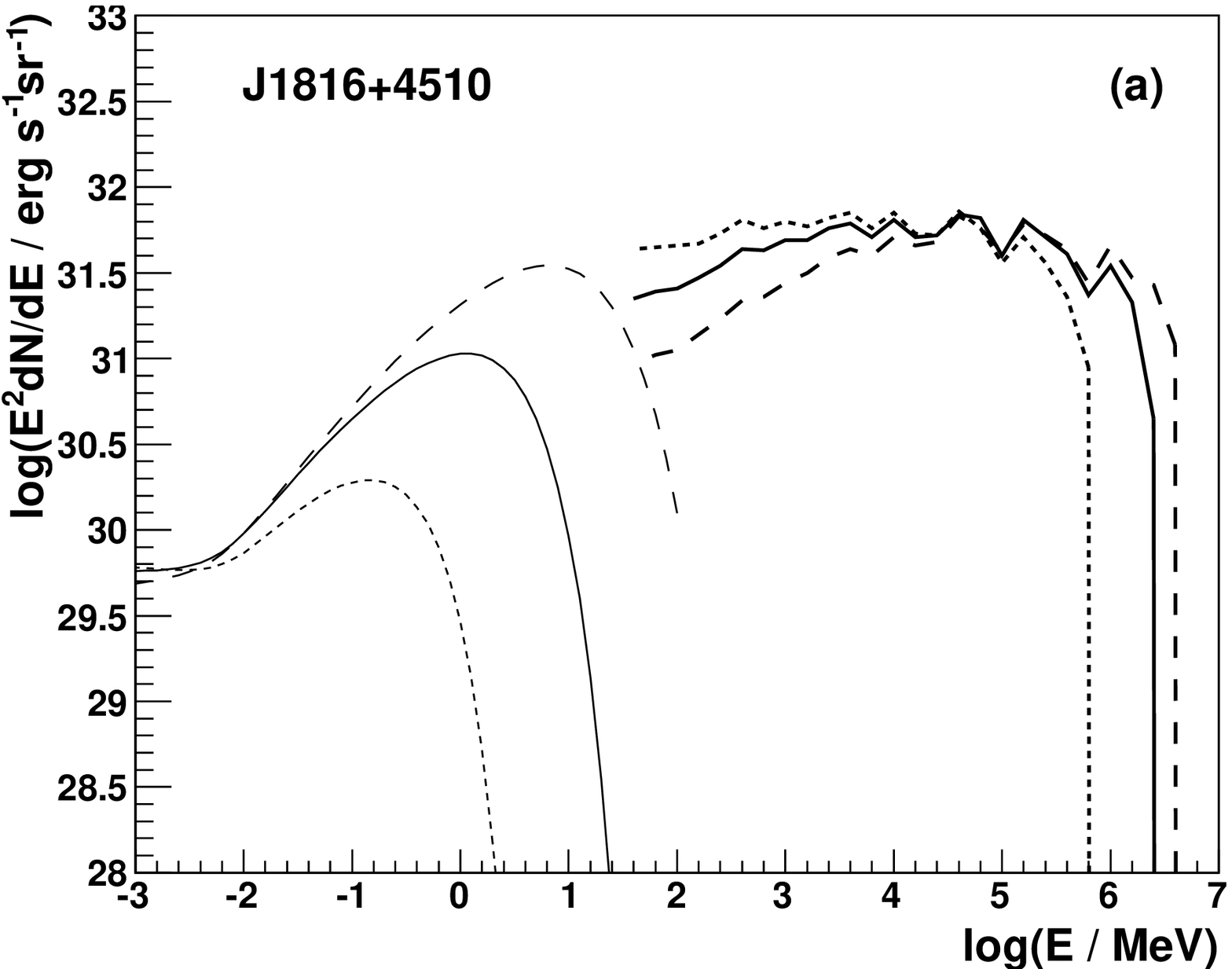}
\includegraphics{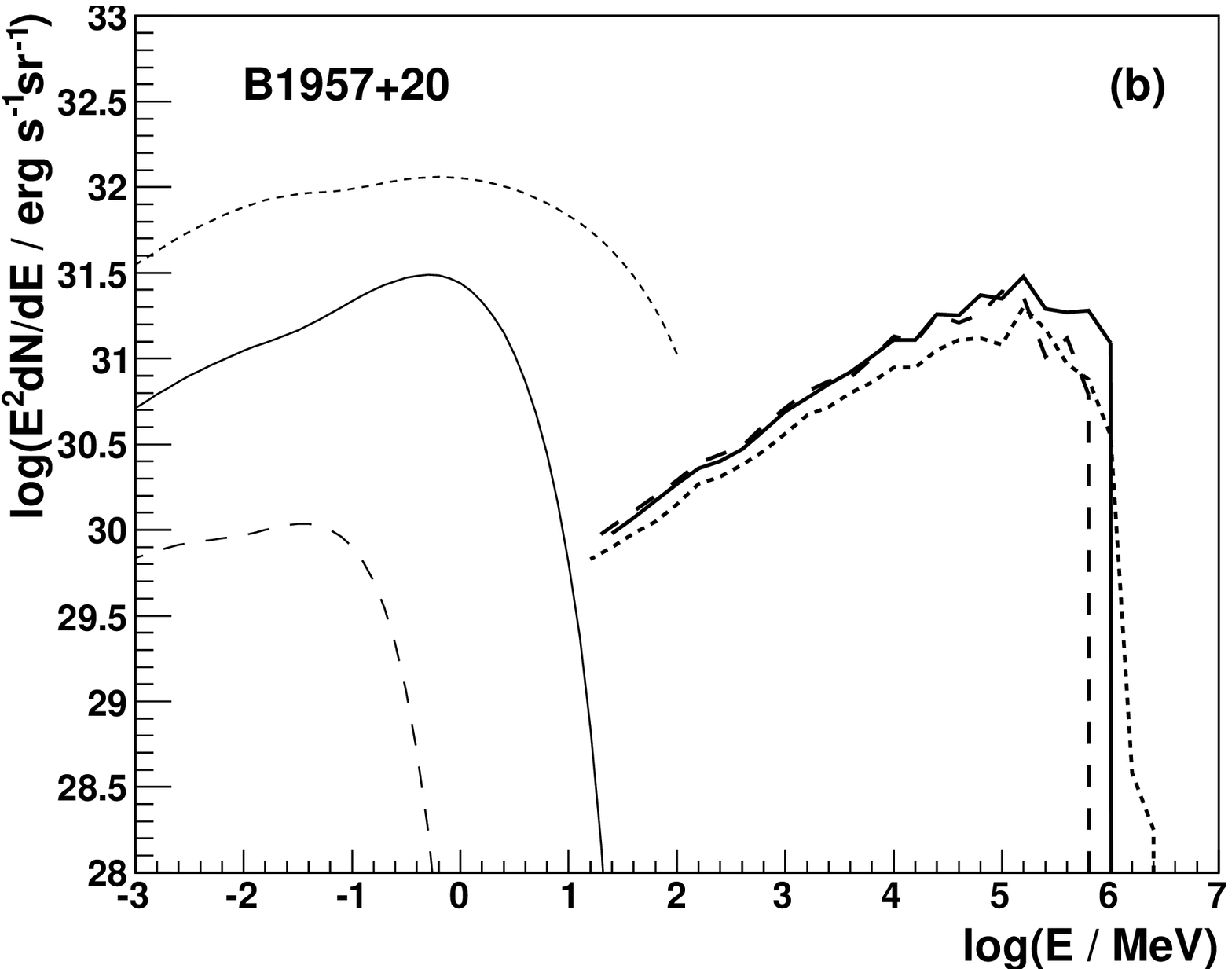}
\includegraphics{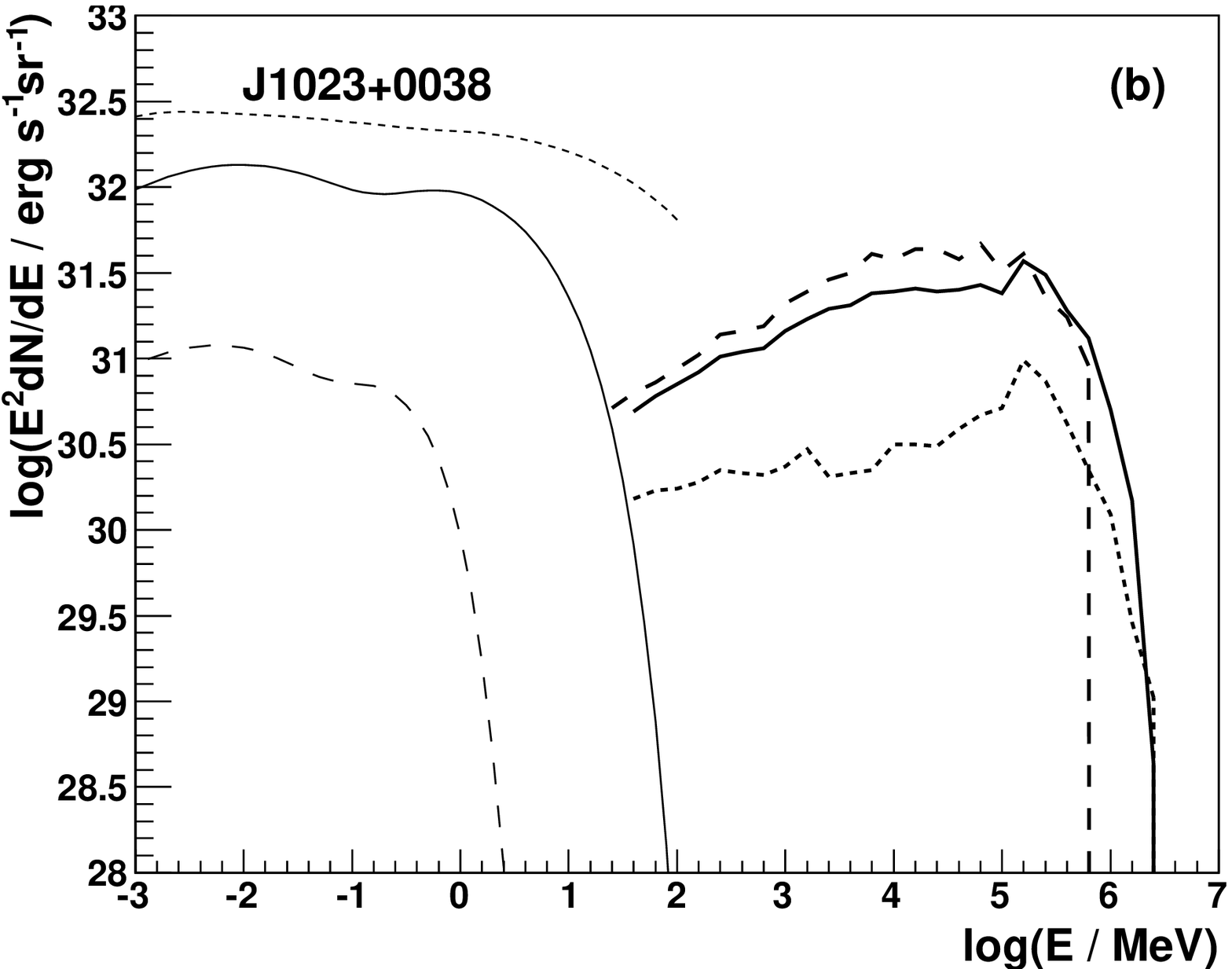}
\includegraphics{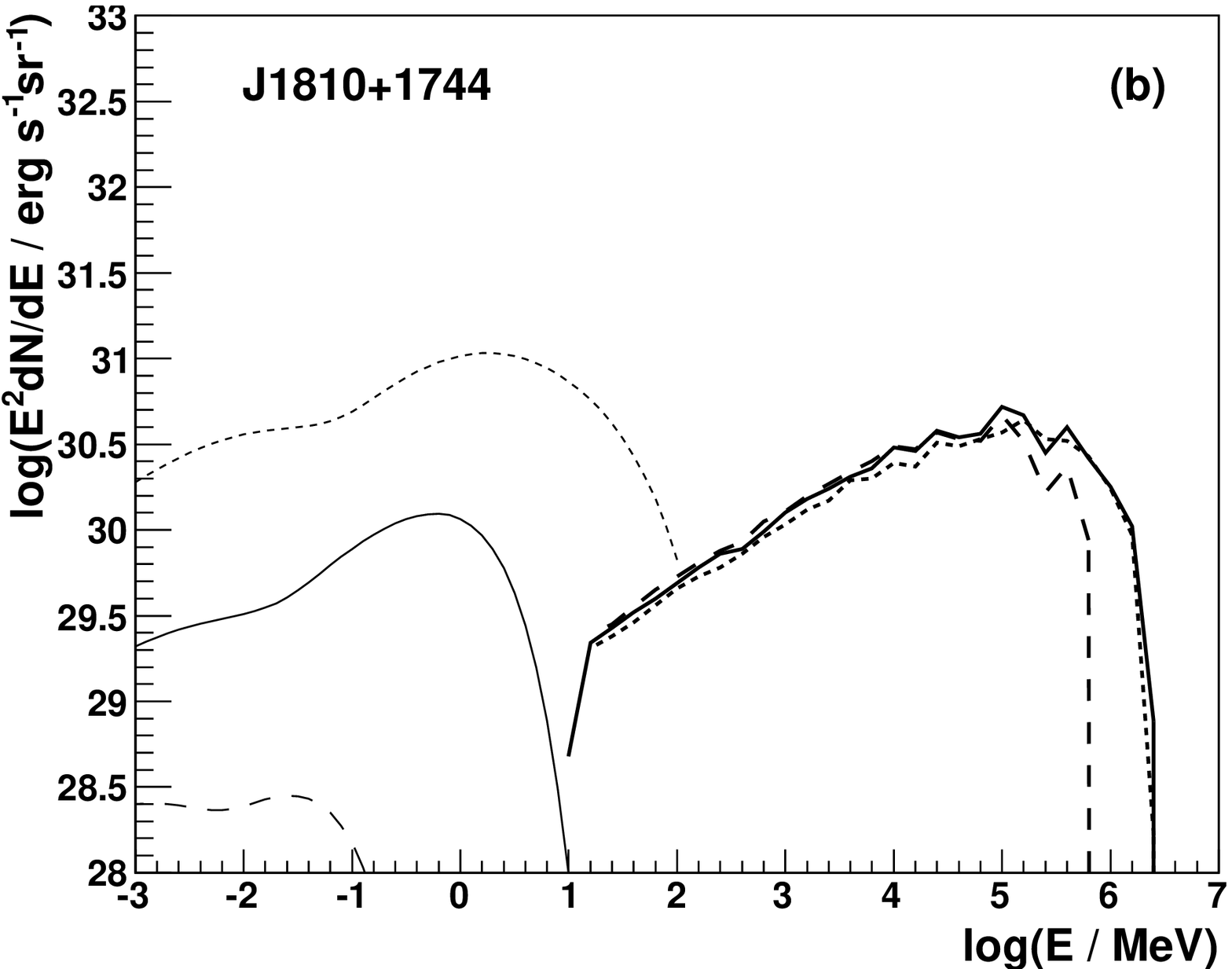}
\includegraphics{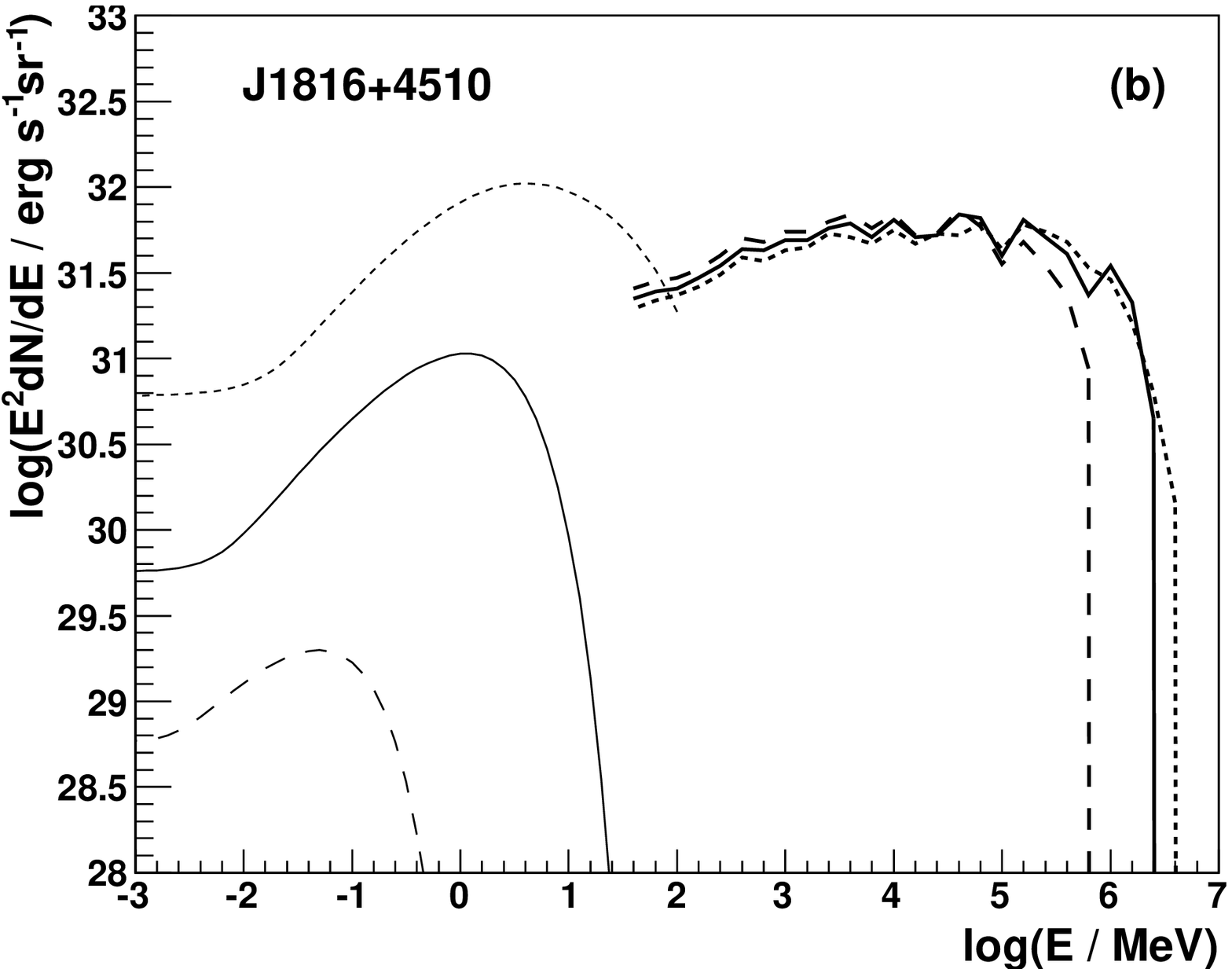}
\includegraphics{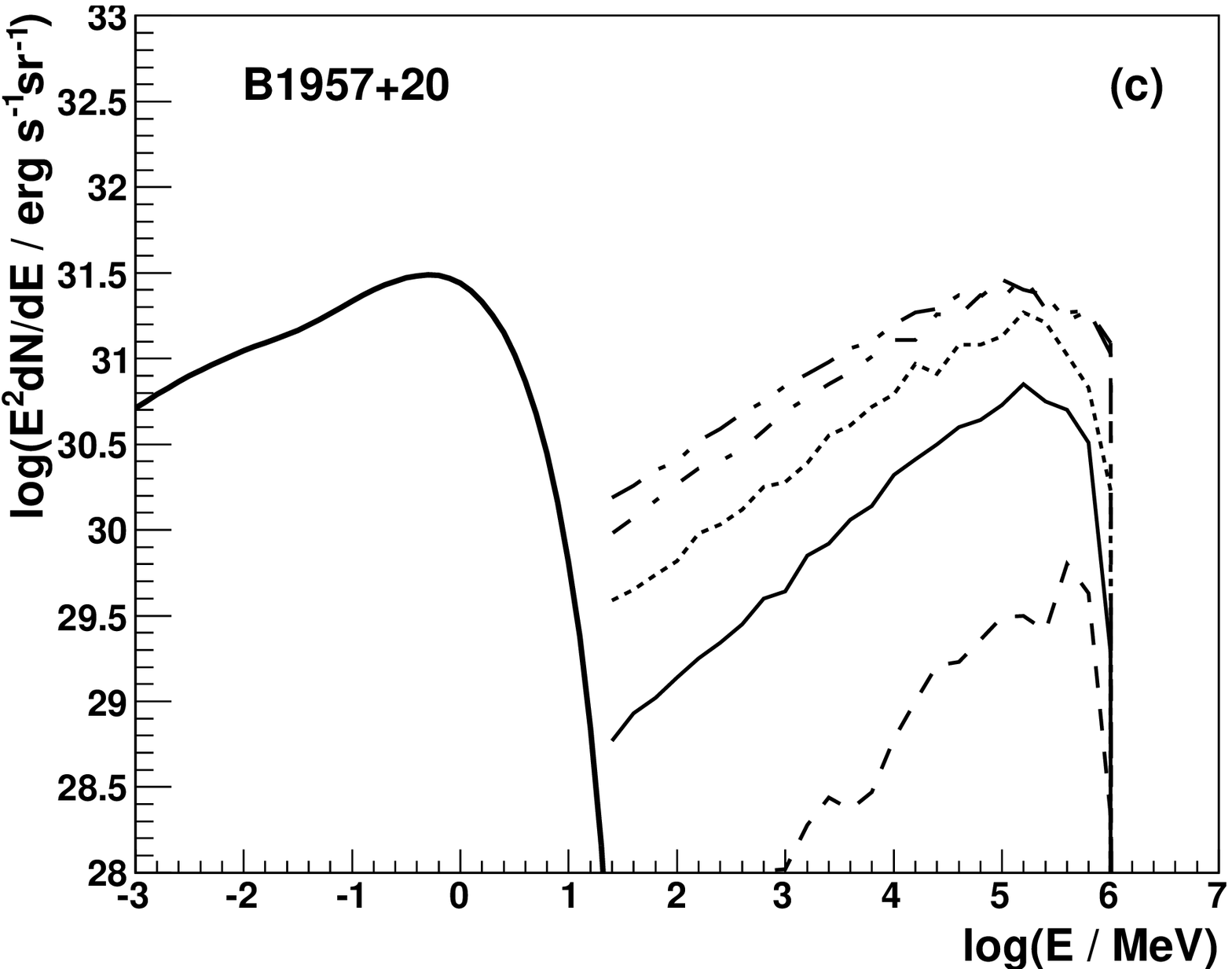}
\includegraphics{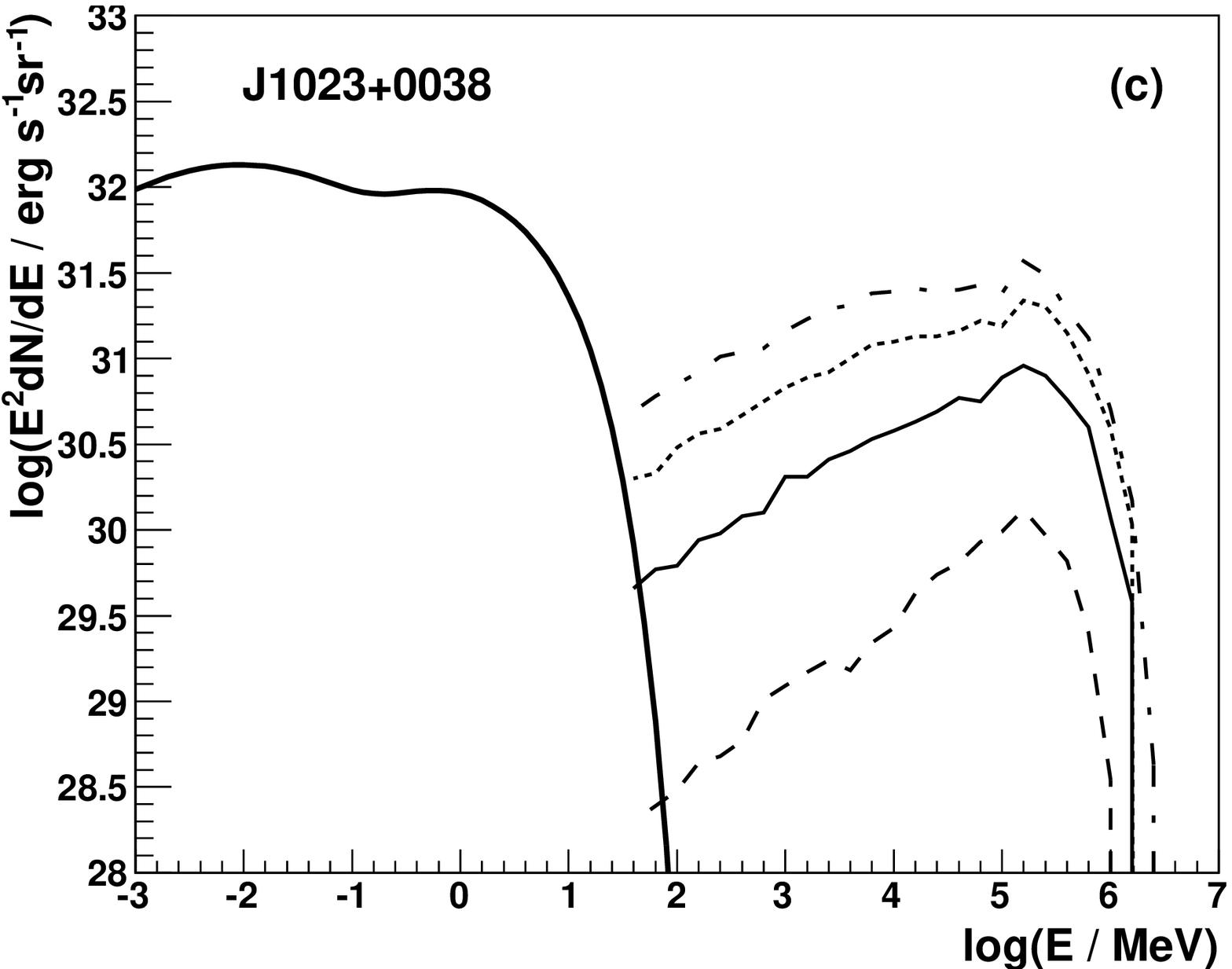}
\includegraphics{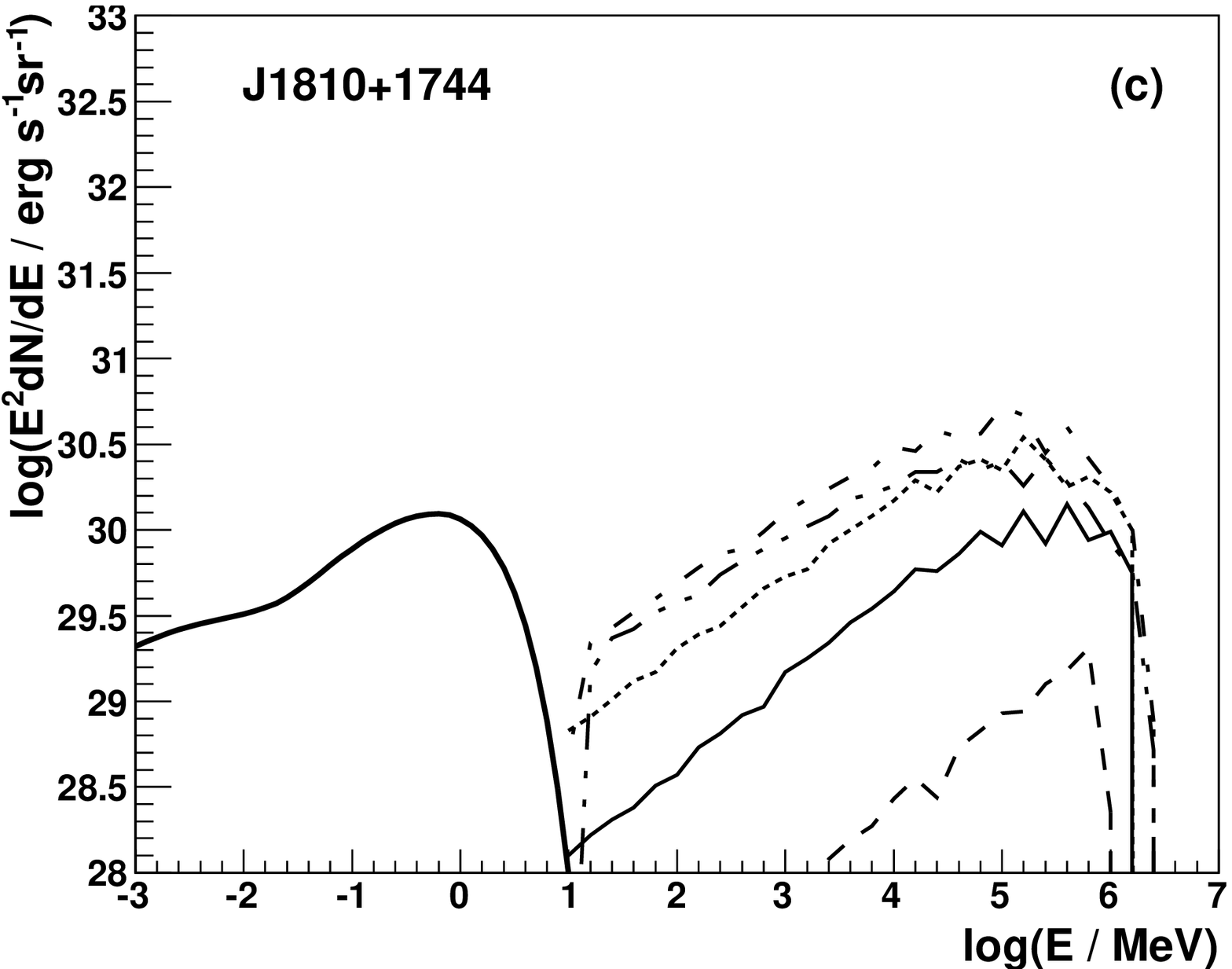}
\includegraphics{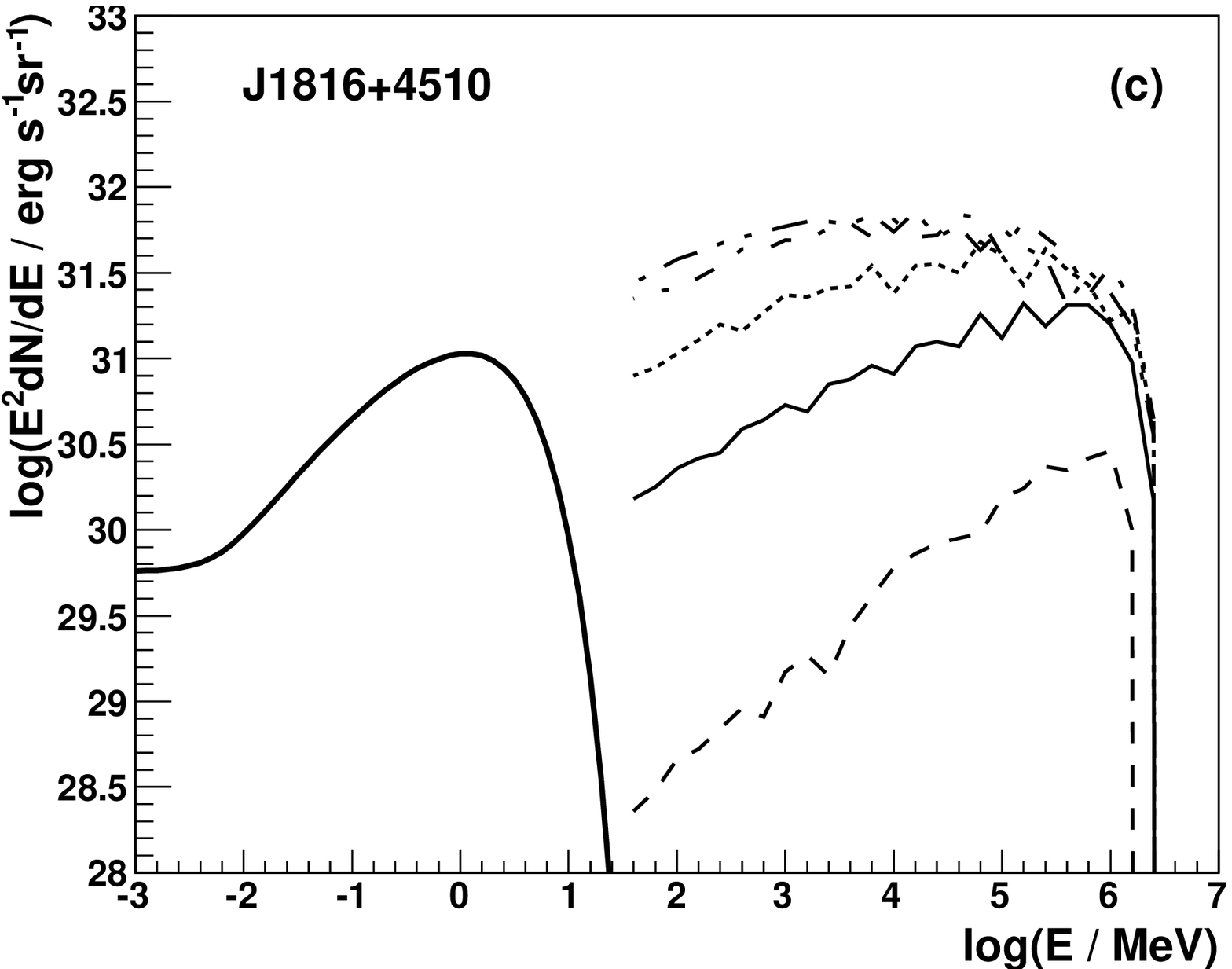}
\caption{Spectral Energy Distribution (energy flux versus energy, SED) of the $\gamma$-rays from IC process and X-rays from synchrotron process, produced by electrons accelerated in the wind collision region in the binary systems: PSR J1816+4510 (right column), PSR J1810+1744 (right-middle column), PSR J1023+0038 (left-middle column), and PSR B1957-20 (left column). 
(a) Dependence on  different advection velocities of the mixed pulsar-stellar winds: $v_{\rm adv} = 10^{10}$ cm s$^{-1}$ (dashed), $3\times 10^9$ cm s$^{-1}$ (solid), and $10^9$ cm s$^{-1}$ (dotted). The other parameters are fixed on: $\sigma = 10^{-3}$, $-0.5 \le \cos\beta \le -0.4$.  The differential spectrum of electrons is assumed to be of the power law type (spectral index -2) between the maximum energy given by Eq.~(7) and minimum energy equal to $E_{\rm min} = 0.5$ GeV.
The power in this electron spectrum is normalized to $1\%$ of the pulsar rotational energy lost by the pulsar.
(b) Dependence on magnetization parameter of the pulsar wind: $\sigma = 10^{-2}$ (dotted), $10^{-3}$ (solid), and $10^{-4}$ (dashed). The other parameters are fixed on: $-0.5 \le \cos \beta \le -0.4$ and $E_{\rm e}$ given by Eq.~7, and $v_{\rm adv} = 3\times 10^{9}$ cm s$^{-1}$.
(c) Dependence on the cosine of the observation angle (measured from the direction defined by the centers of the stars):
$0.9 \le \cos\beta \le 1.0$ (dashed, outwards the companion star), $0.5 \le \cos\beta \le 0.6$ (solid), $-0.1 \le\cos\beta \le 0.$ (dotted), $-0.5 \le \cos\beta \le -0.4$ (dot-dashed), $-1.0 \le \cos\beta \le -0.9$ (dot-dot-dashed), the electron energy $E_{\rm e}$ is given by Eq.~7, $\sigma = 10^{-3}$, and $v_{\rm adv} = 3\times 10^9$ cm s$^{-1}$. 
}
\label{fig2}
\end{figure*}
\section{$\gamma$-ray emission features from specific binaries}

We consider the $\gamma$-ray emission features from four MSP binary systems. Two of them are of the Black Widow type, PSR B1957+20 (Fruchter et al.~1988) and PSR J1810+1744 (Hessels et al.~2011), and two are of the Redback type, PSR J1023+0038 (Archibald et al.~2009) and PSR J1816+4510  (Kaplan et al.~2012). Some basic parameters of these binary systems, important for our calculations, are shown in Table.~1.
The X-ray emission, modulated with the period of the binary system, has been reported from three of these MSP binary systems: PSR B1957+20 (Huang et al.~2012), PSR J1023+0038 (Bogdanov et al.~2011), and PSR J1810+1744 (Gentile et al.~(2013). For PSR J1816+4510, only the upper limit on the X-ray flux has been reported based on the Swift data (Kaplan et al.~2012).
These X-ray positive detections  (and the upper limit) are used farther for the normalization for the synchrotron X-ray emission calculated in terms of our model.
Such procedure allows us to predict the absolute level of the TeV $\gamma$-ray emission expected from the IC scattering of stellar radiation by this same population of relativistic electrons. 
In this calculations we fix the injection place of relativistic electrons between the pulsar and the companion star at the distance of the collision region from the companion star (which is also reported in Table.~1).

At first, we investigate how the synchrotron and IC $\gamma$-ray spectra depend on the free parameters of the considered model for all four binary systems. In Fig. 2a we show how the spectra depend on the velocity of the mixed pulsar-stellar winds considering the range between $v_{\rm adv} = v_{\rm mix} = 10^9 - 10^{10}$ cm s$^{-1}$. 
This range of velocities of the mixed pulsar-stellar winds is expected for the typical parameters of the MSP binary systems, i.e. pulsar energy loss rate $L_{35} = 1$, the mass loss rate of the companion star $\dot{M}_{-11} = 1$, and the solid angle over-taken by the collision region $\xi_{-1} = 1$  (see Eq.~1).
Another important parameter, which determine the model, is the magnetization parameter of the pulsar wind. We apply the value for $\sigma = 10^{-3}$. It is of the order of that derived for the pulsar wind termination shock in the Crab Nebula ($\sim 0.003$, Kennel \& Coroniti 1984), but clearly below this value since we expect that the recconnection of the magnetic field in the collision region of the turbulent stellar wind and the pulsar wind will occur more efficiently within the MSP binary systems. 
We also fix the range of cosine of observation angles on $-0.5 \le \cos\beta \le -0.4$, since for such range of angles the largest $\gamma$-ray fluxes are expected from the MSP binary systems.
Electrons are injected with the differential power law spectrum (spectral index -2) at energy range between the maximum energy given by Eq.~(7)  and the minimum energy $E_{\rm min} = 0.5$ GeV. For the illustration purposes we normalize the power in this electron spectrum to $1\%$ of the rotational energy lost by the pulsar.
This normalization coefficient is the product of the solid angle overcomed by the colliding winds and the efficiency of electron acceleration in turbulent region.
Note that the shock wave, formed in the interaction of the pulsar and stellar winds, is expected  to overcome typically about $\sim$10$\%$ of the whole solid angle
(for the parameters reported in Table~1). On the other hand, the efficiency of particle acceleration in the pulsar shock is expected to be of the order of 
$\sim$10-30$\%$ (e.g. nebula around the Crab pulsar). In fact, this coefficient is of the order of magnitude of that derived farther in Section 5.1, from the normalization of synchrotron spectra, calculated in terms of our model, to the X-ray observations of the considered millisecond pulsar binary systems.

Note that only the lower energy part of the IC $\gamma$-ray spectrum strongly dependents on the velocity of the mixed winds. For faster winds, the $\gamma$-ray emission below $\sim$100 GeV is suppressed since electrons do not spend enough time close to the companion star in order to cool completely. The higher energy part of the IC spectrum is not so sensitive on the maximum energy of electrons due to the importance of the synchrotron process.
However, the synchrotron spectra for faster winds peak at larger energies since the maximum energies of electrons accelerated in the transition region strongly depend on the mixed wind velocity (see Eq.~7).

On Figs.~2b we show the dependence of the synchrotron and IC spectra on the magnetization parameter of the pulsar wind which, in our model, determines the strength of the magnetic field at the transition region. We fix in this case: the mixed wind velocity on $3\times 10^9$ cm s$^{-1}$, the maximum energy of electrons by applying Eq.~(7), and the range of the cosine of observation angles on $-0.5 \le \cos\beta \le -0.4$. The synchrotron spectra are very sensitive on the magnetization parameter $\sigma$ (flux, maximum energies). On the other hand, the IC $\gamma$-ray spectra only weakly depend on the magnetization of the mixed wind, except for the case of the PSR J1023+0038 binary system in which the radiation field created by the companion star is the weakest among all considered binaries. 

Finally, in Figs.~2c we show the dependence of the IC $\gamma$-ray spectra on the range of the observation angles. The peak in the $\gamma$-ray spectra shifts slowly to higher energies but the flux increases fast for larger angles $\beta$. Both effects
are due to more efficient cooling of relativistic electrons directed towards the companion star. These directions correspond to the lower values of the cosine of the observation angles
(i.e. larger values of the angles $\beta$). Therefore, our model predicts clearly stronger IC $\gamma$-ray production at phases when the pulsar is behind the companion star.

\subsection{Expected gamma-ray spectra}

\begin{figure*}[t]
\vskip 10.truecm
\includegraphics{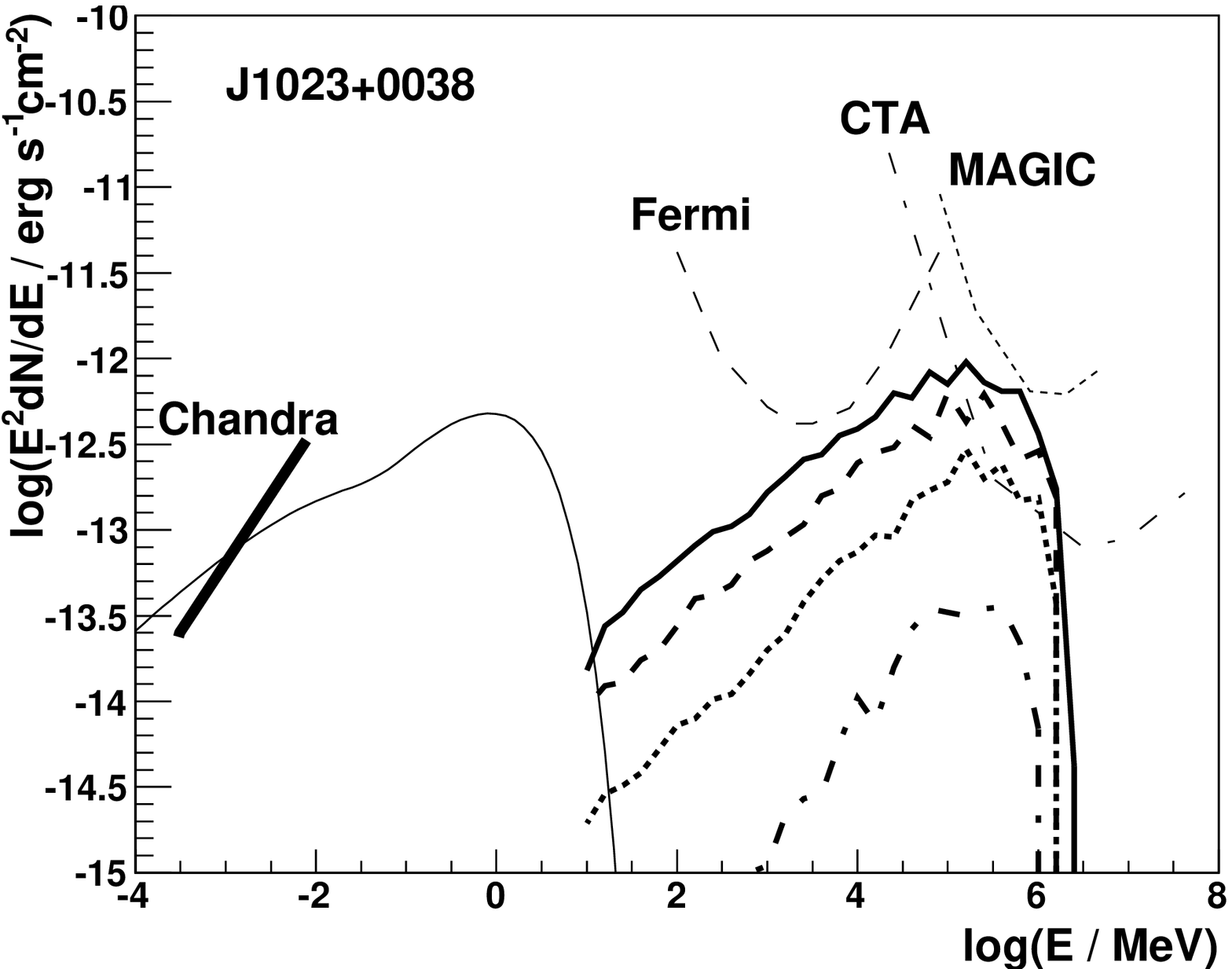}
\includegraphics{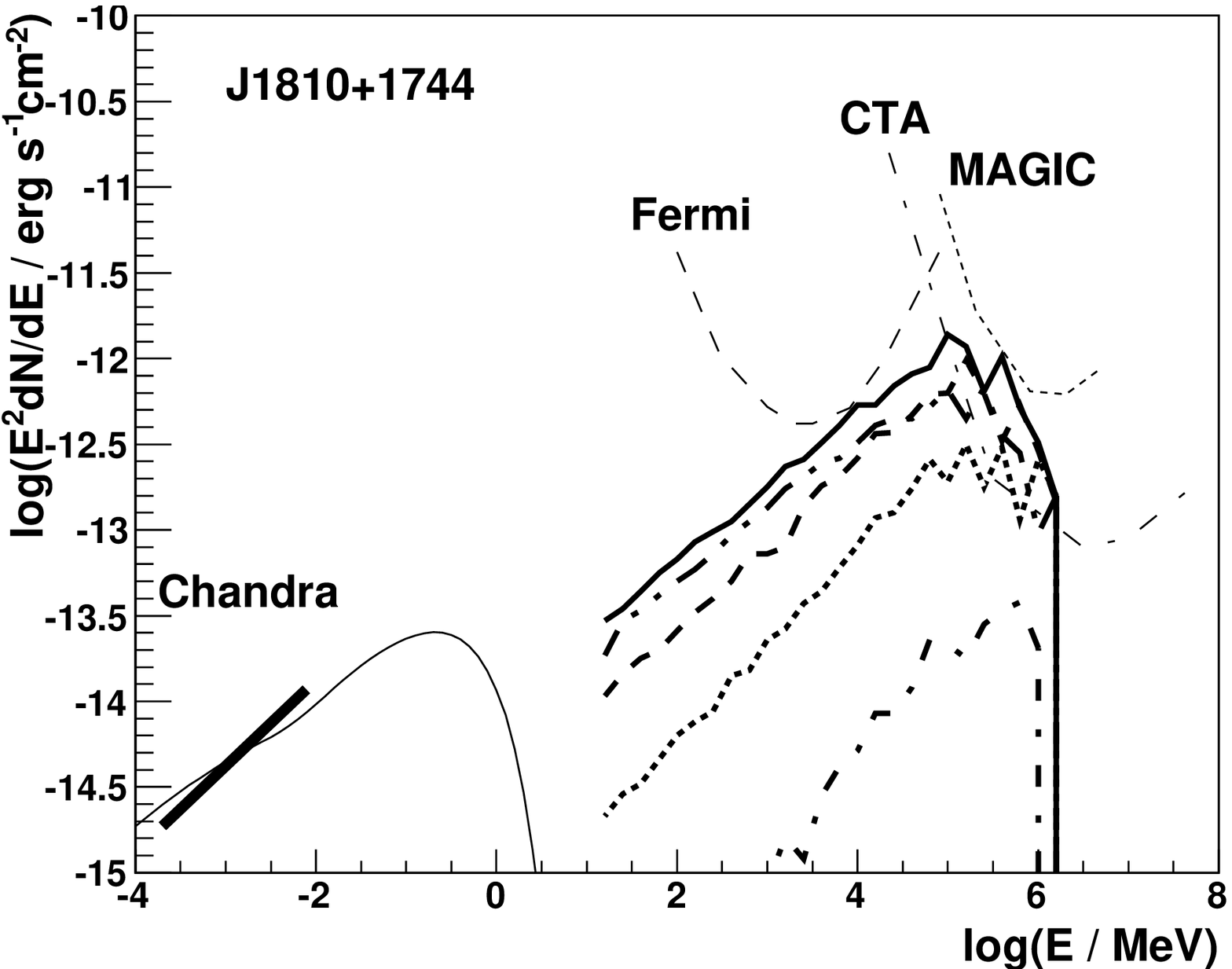}
\includegraphics{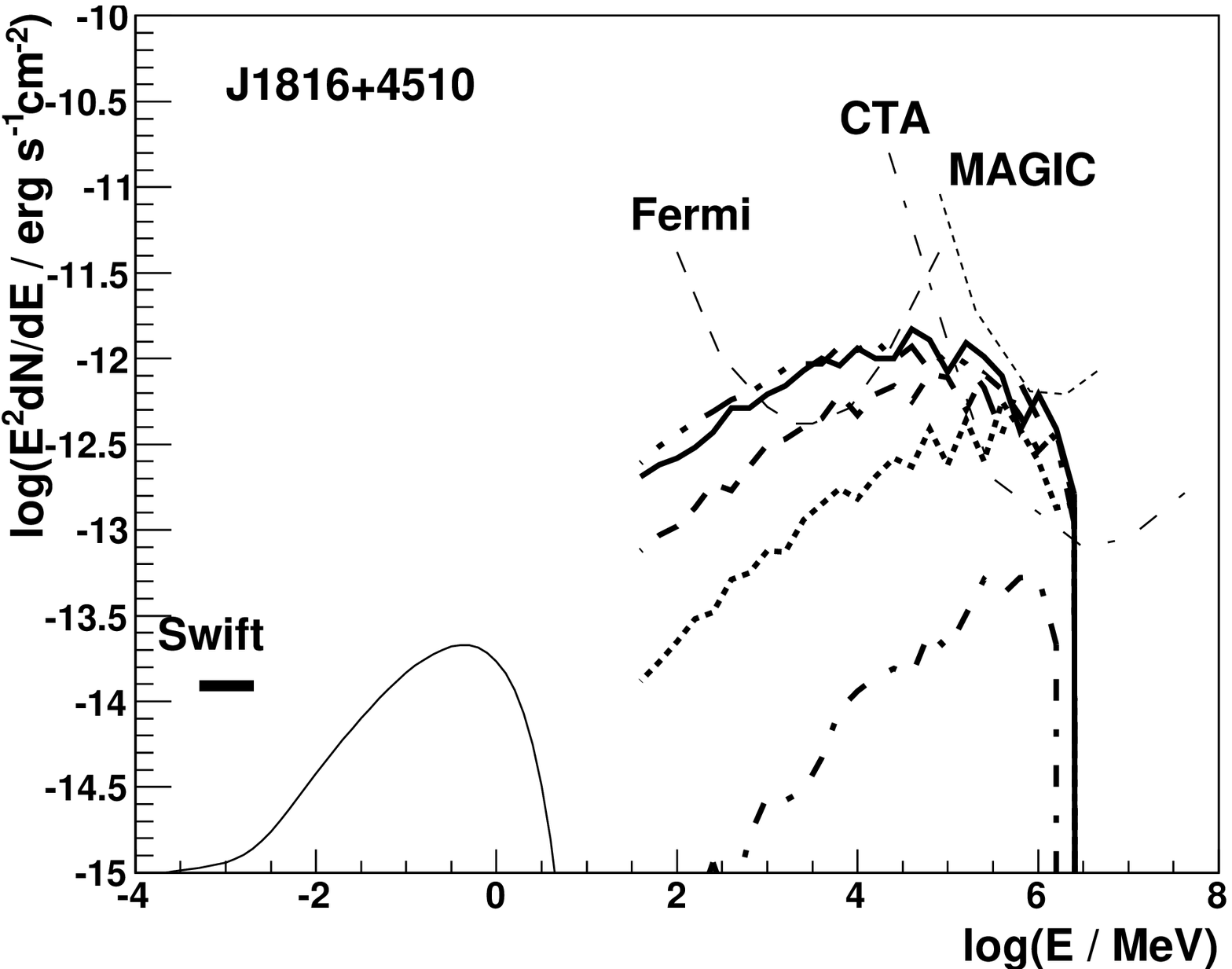}
\includegraphics{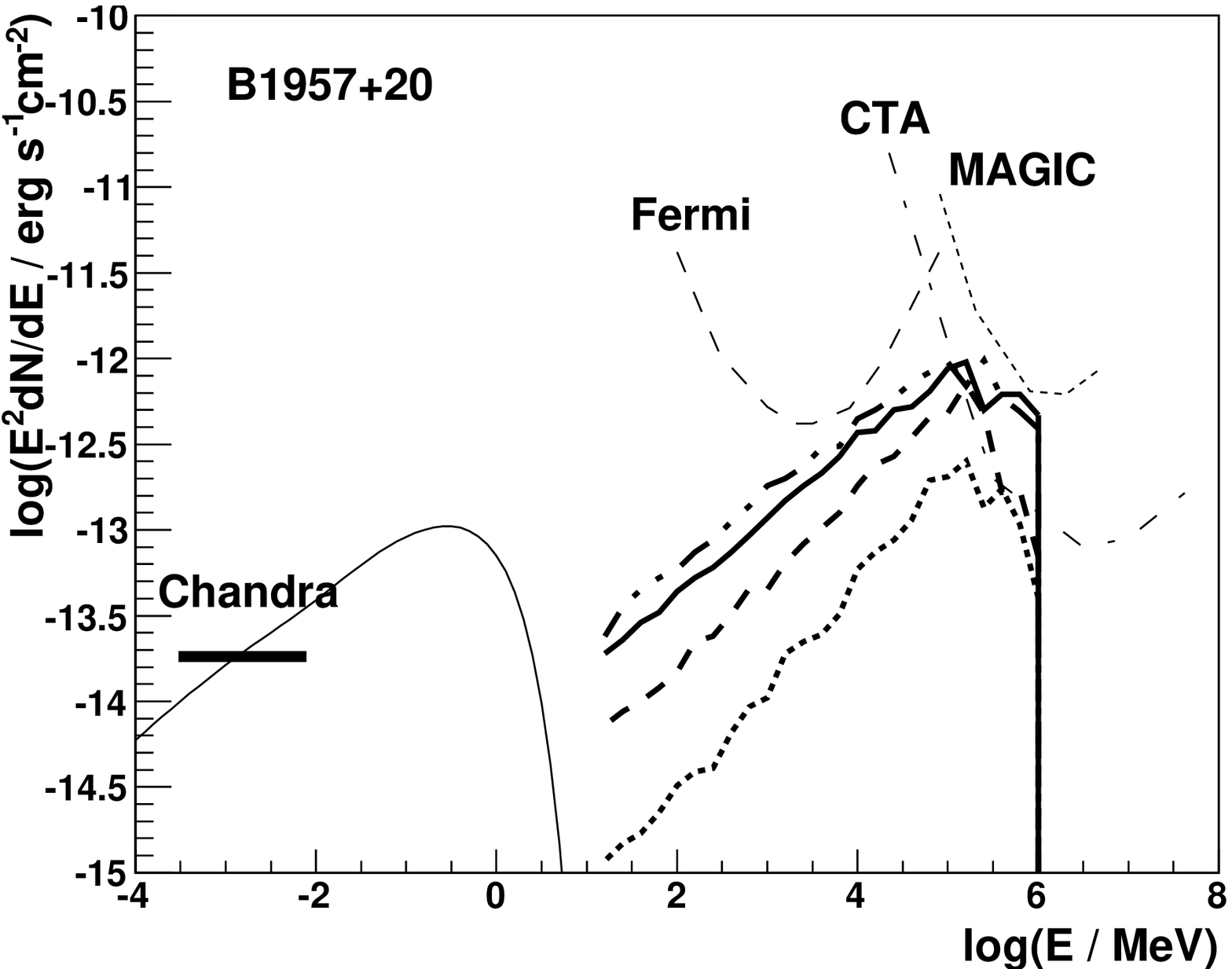}
\caption{Comparison of the high energy emission  (energy flux versus energy) expected from the MSP binary systems with the sensitivities of Fermi-LAT (10 yrs - extragalactic background, see thin dashed curve, Funk et al.~2013), MAGIC (thin dotted curve, Aleksi\'c et al.~2012) and CTA (thin dot-dashed, Bernl\"ohr et al.~2012) for different range of observation angles: $0.9 \le \cos\beta \le 1.0$ (dot-dashed, outward the companion star), $0.5 \le \cos\beta \le 0.6$(dotted), $-0.1 \le \cos\beta \le 0.$ (dashed), $-0.5 \le \cos\beta \le -0.4$ (solid), $-1.0 \le \cos\beta \le -0.9$ (dot-dot-dashed). The other parameters of the models have been fixed on: $\sigma = 10^{-4}$, and $v_{\rm adv} = 10^{10}$ cm s$^{-1}$, and $E_{\rm e}$ is given by Eq.~7 . The spectra are normalized to the X-ray fluxes observed from PSR B1957+20 (Huang et al.~2012), PSR J1023+0038 (Bogdanov et al.~2011), PSR J1810+1744 (Gentile et al.~2013). In the case of PSR J1816+4510 the spectrum is consistent with the upper limit derived from the Swift data (Kaplan et al.~2012). This normalization requires the energy conversion efficiency from pulsars to relativistic electrons estimated on: $1.6\times 10^{-2}$ (for PSR J1023+0038), $6.5\times 10^{-2}$ (PSR J1810+1744), $5\times 10^{-3}$ (PSR J1816+4510), and $1.8\times 10^{-3}$ (PSR B1957+20).}
\label{fig3}
\end{figure*}

The synchrotron and IC $\gamma$-ray spectra, calculated for likely parameters defining considered model, are confronted with the observations of the modulated synchrotron emission in the X-ray energy range (or with derived upper limit) reported in different works.
The constraints on the modulated X-ray emission from these four binary systems are reported in Table.~1.
We apply the velocity of the mixed pulsar-stellar wind equal to $10^{10}$ cm s$^{-1}$, which is of the order of that estimated in Section.~2. For the assumed magnetization parameter, $\sigma = 10^{-4}$, and the other typical parameters of the pulsar and its binary system, we estimate the maximum energies of accelerated electrons by using Eq.~7. Note that the assumed value of the magnetization parameter is clearly lower than estimated in the case of the Crab Nebula
($\sim$0.003, Kennel \& Coroniti~1984). However, in our model the pulsar wind is expected to mix very efficiently with the inhomogeneous stellar wind in the transition region. Then, due to the magnetic recconnection process, a relatively weak magnetic field is expected in the acceleration and radiation region. In Fig.~3 we show the expected IC $\gamma$-ray spectra produced by electrons with the differential power law spectrum (spectral index -2), after normalization to the observed modulated X-ray emission from binary systems, PSR J1023+0038, PSR J1810+1744, and PSR B1957+20, and consistent with the upper limit derived for the binary system J1816+4510. Specific curves show the results for different ranges of the cosine of the observation angles, i.e. related to different  phases of the binary system. The IC $\gamma$-ray spectra in the sub-TeV energy range are on the level of sensitivity of the present Cherenkov telescopes (such as HESS, MAGIC and VERITAS), for the observation angles which are optimal for efficient $\gamma$-ray production (i.e. pulsar behind the companion star).
 
In the case of three binary systems, PSR B1957+20, PSR J1023+0038 and PSR J1810+1744, the estimates of the inclination angles are available in the literature: $65^\circ$ for PSR B1957+20 (Reynolds et al.~2007) and $46^\circ$ for PSR J1023+0038 (assuming that the neutron star mass is equal to the $1.4$ Solar masses, Archibald et al.~2009, Archibald et al.~2010), and $48^\circ$ for PSR J1810+1744. Note, that the values of these angles depend on the masses of the neutron stars in these binary systems which are in fact not well known.
The values of the inclination angles determine the lowest values of the cosine of the angle $\beta$, for the pulsar and the observer located on the opposite site of the companion star 
(i.e. $\omega = \pi$). Then, $\cos\beta$ equals to $-0.9$ (for PSR B1957+20), $-0.72$ (for PSR J1023+0038) and $-0.74$ (for PSR J1810+1744).
It is clear from Fig.~3 that for these locations of pulsars on the orbits, the TeV $\gamma$-ray fluxes are expected to be close to the maximum values. However, when the pulsar is in front of the companion star, the fluxes should be a factor of $2-3$ lower in the TeV $\gamma$-ray energy range and about an order of magnitude lower in the GeV $\gamma$-ray energies (see Fig.~3). 

We conclude that the TeV $\gamma$-ray signals from these binary systems have a chance to be detected in the extensive observations with the present Cherenkov telescopes only during specific range of orbital phases. 
The situation should be much better with the construction of the next generation of Cherenkov telescope arrays such as CTA. In the case of CTA, the TeV $\gamma$-ray emission from MSP binary systems is expected to be probed for the most of the range of orbital phases.
Note that modulated GeV $\gamma$-ray emission from the considered binary systems is not likely to be detected in the Fermi-LAT data. The exception can only be the Redback type binary, PSR J1816+4510, which contains exceptionally luminous companion star with the surface temperature above $\sim$10$^4$ K (Kaplan et al.~2013). Recently identified $\gamma$-ray source in the direction of this pulsar show clear pulsations with the millisecond pulsar period (Kaplan et al.~2012). Interestingly, the $\gamma$-ray spectrum calculated for PSR J1816+4510 seems to be atypical for the millisecond pulsar population. It shows emission above the characteristic cut-off in the pulsar's spectra at $\sim$2-3 GeV. It might be that a part of this emission comes from the binary system itself. This interesting  binary system should be clearly studied deeper in the future since positive detection (or even the upper limits) of the $\gamma$-ray component modulated with the binary period will provide important constraints for the scenario discussed in this paper.

\subsection{Gamma-ray light curves}

\begin{figure*}
\vskip 6.truecm
\includegraphics{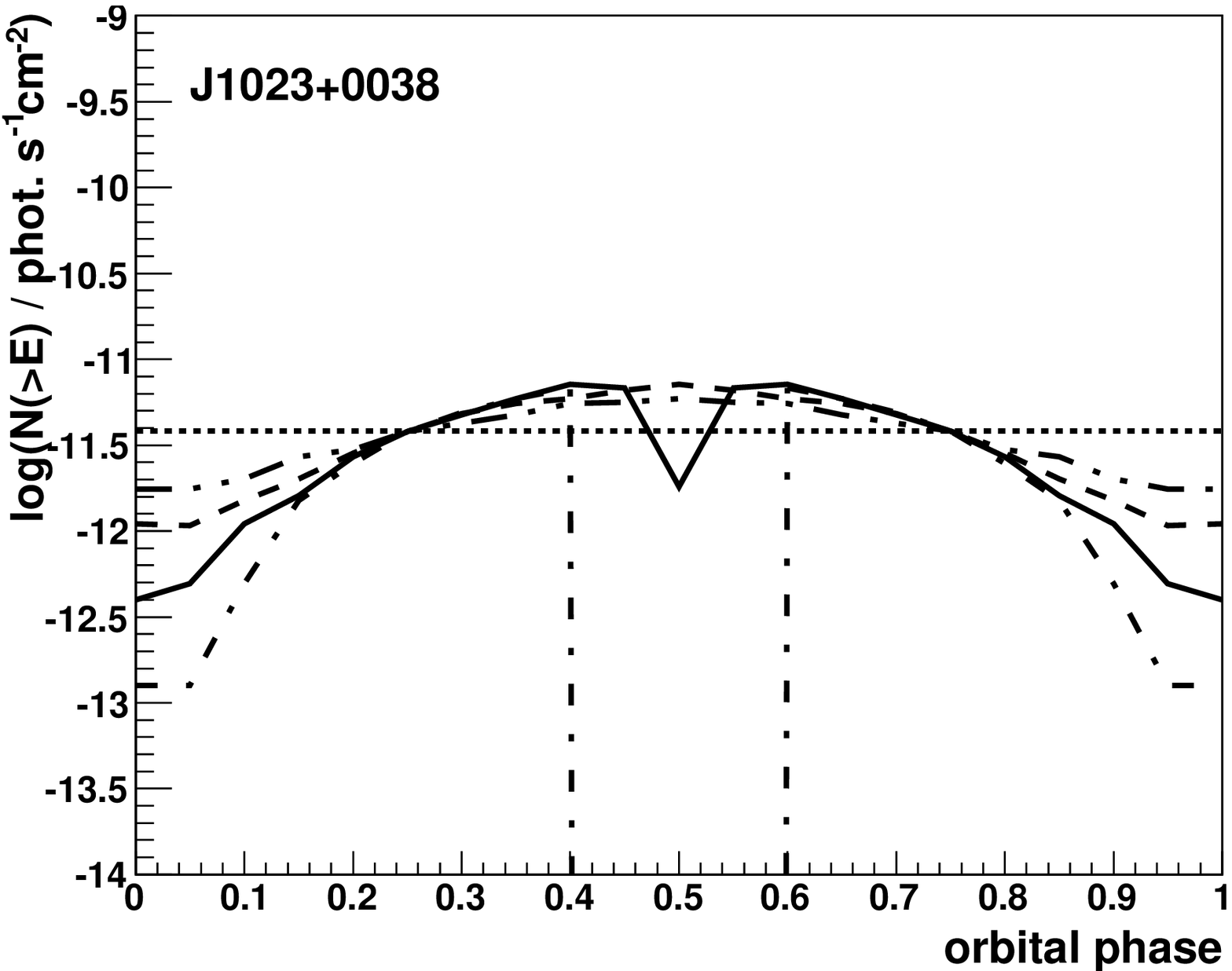}
\includegraphics{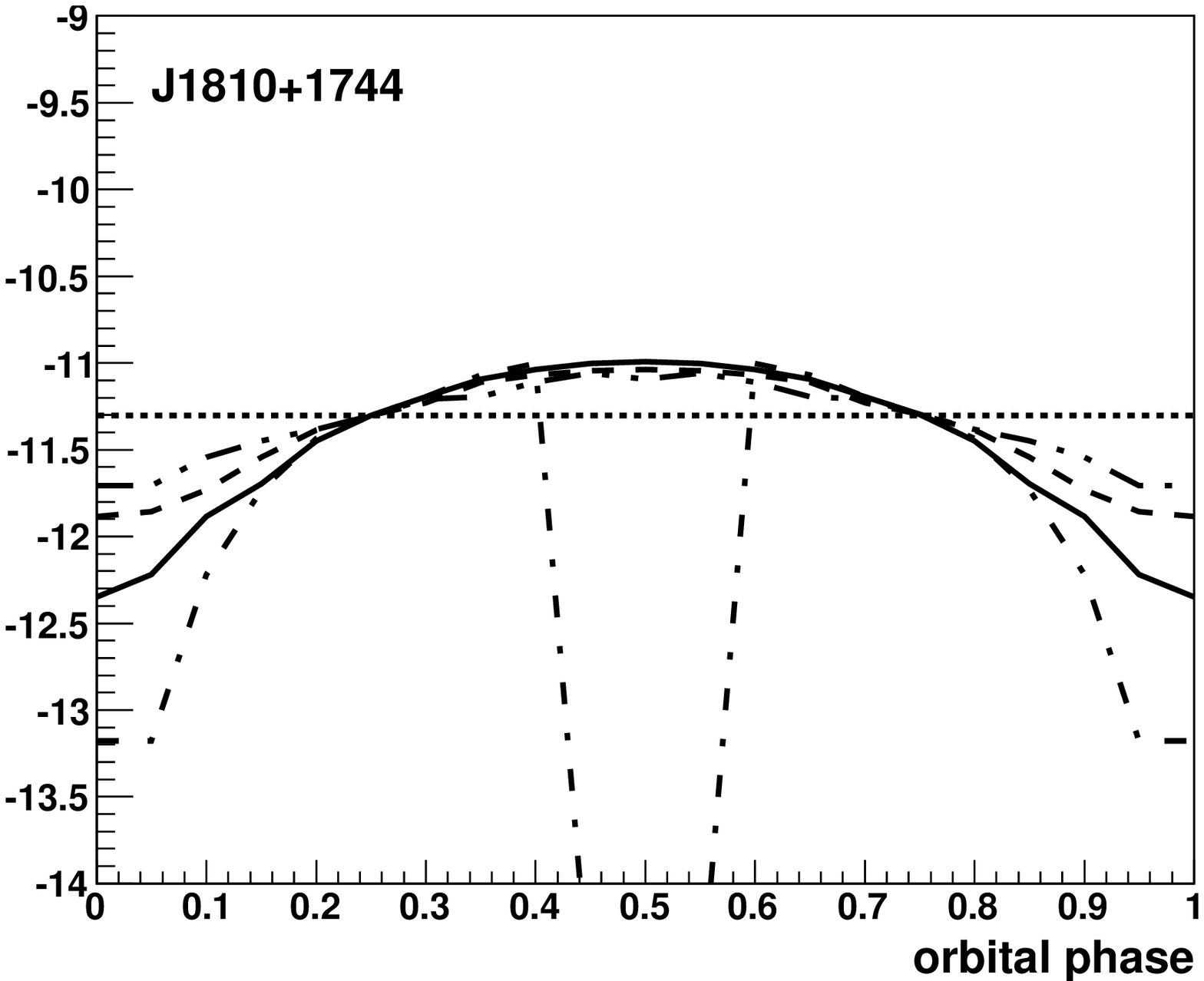}
\includegraphics{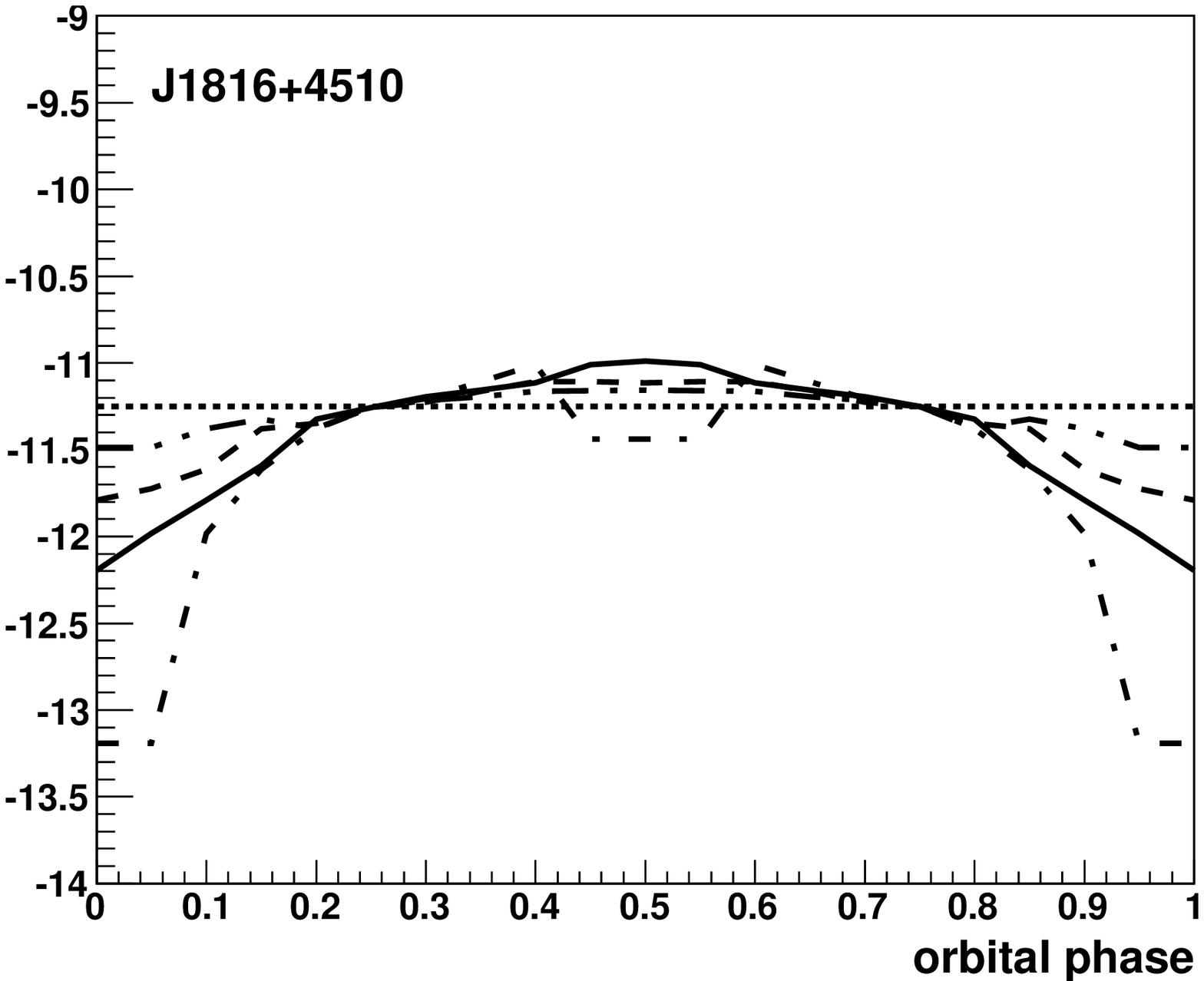}
\includegraphics{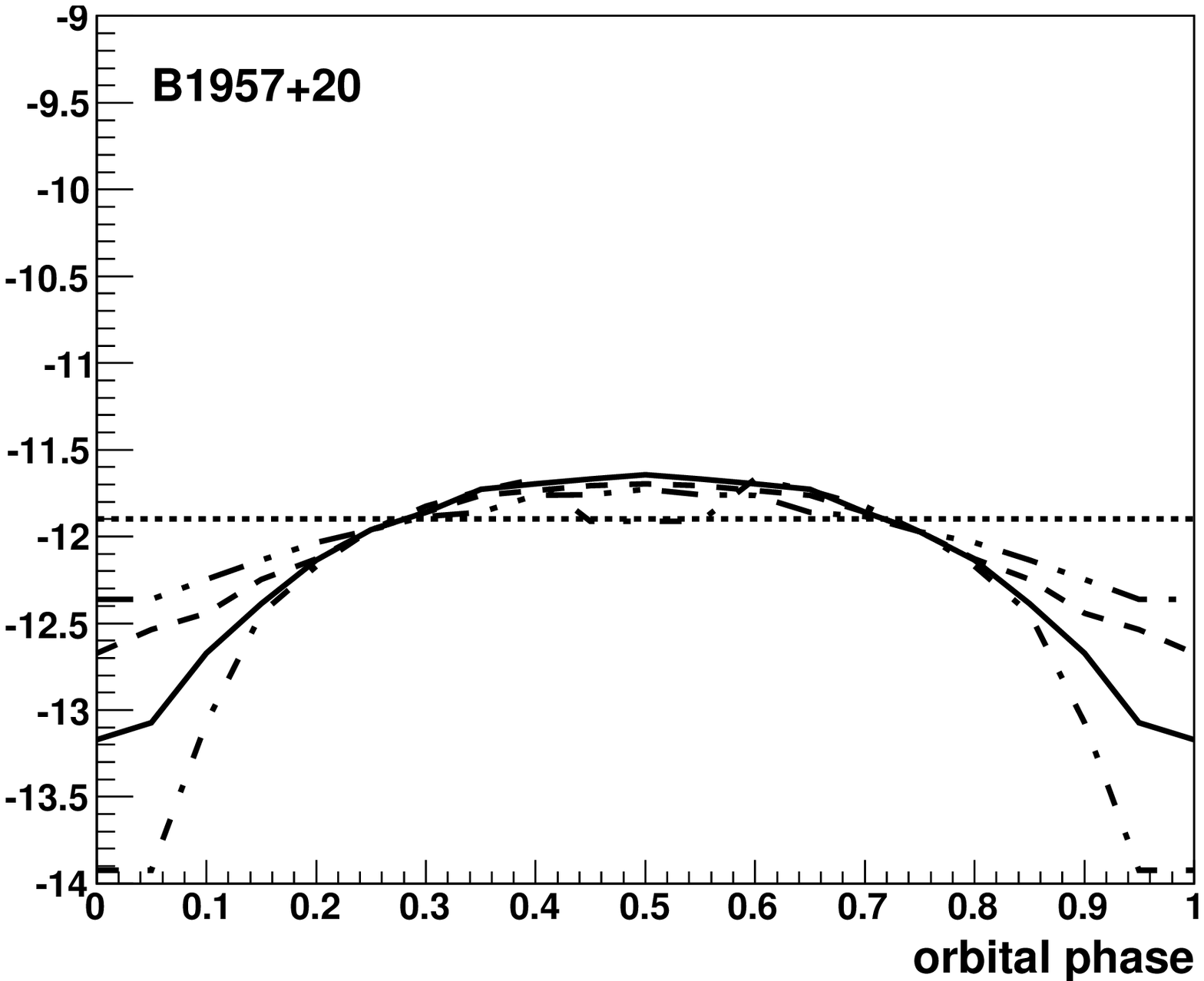}
\caption{The $\gamma$-ray light curves in the TeV energies ($>$100 GeV)
expected from considered MSP binary systems: PSR J1023+0038 (on the left), PSR J1810+1744 (left-middle), PSR J1816+4510 (right-middle), and PSR B1957+20 (right). Specific curves show the results for different inclination angles of the binary systems: $i = 0^\circ$ (dotted), $30^\circ$ (dot-dot-dashed), $45^\circ$ (dashed), $60^\circ$ (solid), and $90^\circ$ (dot-dashed). The phase is counted from the location of the MSP in front of the companion star. The parameters of the companion stars and the binary systems are reported in Table.~1. It is assumed that the MSP moves in a circular orbit around the companion star. The parameters describing the spectrum and escape of electrons are the same as in Fig.~3. The injection place of electrons is assumed to be at the apex of the collision region (described by $R_{\rm sh}$ in Table~1). The $\gamma$-ray fluxes are collected in the range of phases with the width equal to 0.05.}
\label{fig4}
\end{figure*}

In order to give more quantitative predictions, which could be tested by the extensive observations in the TeV $\gamma$-ray energy range, we calculate the expected $\gamma$-ray light curves at energies above $>$100 GeV from all four MSP binaries (see Fig.~4). 
The phase zero is measured from the location of the pulsar on the orbit when it is in front of the companion star. Different curve styles indicate the cases with the observer located at different inclination angles starting from $i = 0^\circ$ (dotted) up to $90^\circ$ (dot-dashed). 

For the large inclination angles, the light curves show clear dip (around the phase 0.5), which is due to shadowing effect by the companion star. The extent of the dip depends on the relative separation of the stars in the binary system (in respect to the stellar radius).
Note also that the TeV $\gamma$-ray flux depends rather weakly on the inclination angle of the binary system for the range of phases when the pulsar is behind the companion star. On the other hand, we notice strong change of the TeV flux with the inclination angle for the phases when the pulsar is in front of the star. This behaviour is due to the basic feature of the IC scattering process in the case of strongly anisotropic radiation field. The efficiency of IC scattering depends much stronger on the 
collision angle between electron and soft photon when the photon arrives from the hemisphere behind the direction of the electron. 

Let us discuss in detail the case of three MSP binaries for which the inclination angles are known.
In the case of the Redback MSP binary, PSR J1023+0038, and the Black Widow binary, PSR J1810+1744, we predict modulation of the TeV $\gamma$-ray signal with the orbital period of the binary system by a factor of $\sim$3-5. However, in the case of the Black Widow binary system, PSR B1957+20, the predicted modulation of the signal is clearly over one order of magnitude. 
Therefore, it is so important to observe MSP binary systems at the correct phase intervals in order to guarantee  positive detection. The stronger modulation pattern in the case of PSR B1957+20 is due to the larger inclination angle and weaker radiation field created by the companion star in this binary system in respect to the Redback binary PSR J1023+0038.

The TeV $\gamma$-ray flux is expected to be the most sensitive on the inclination angle of the binary system when the pulsar is in front of the companion star (corresponding to phase zero, see Fig.~4). In principle, CTA upper limits in this phase, or positive detection which will be rather difficult due to the low fluxes, should allow to put the limit on the inclination angle of the binary system. However, this limit may not be very restrictive due to relatively low TeV $\gamma$-ray fluxes at these phases, comparable the sensitivity of CTA.

The TeV $\gamma$-ray light curves, predicted by our model for the MSP binary systems, show basically different emission features than observed from the massive TeV $\gamma$-ray binary systems such as LS~5039 (Aharonian et al.~2006) or LSI~303~+61 (Albert et al.~2009). 
The massive binaries 
show the maximum of the TeV emission at phases when the compact objects are generally in front of the massive stars. In contrast, we predict here that the TeV $\gamma$-ray light curves from MSP binaries should behave completely opposite, i.e. the maximum of the TeV emission should occur at phases shifted by half of the period in respect to massive TeV binaries. 
Therefore, observations of MSP binary systems with Cherenkov telescopes should concentrate
on phases when the pulsar is behind the companion star in order to increase the chances for
detection of the TeV $\gamma$-ray signal.

\section{Discussion and Conclusion}

We consider the high energy processes which are expected to occur within the MSP binary systems of the Redback and Black Widow types. We modify the general scenario proposed for such objects (acceleration of particles and their radiation in the region of the colliding pulsar and companion star winds, e.g. Harding \& Gaisser~1990, Arons \& Tavani~1993) by assuming that the wind of the companion star, induced by the energy realised by the pulsar, is very inhomogeneous. This assumption has two 
important consequences. Firstly, winds should mix efficiently at the collision region and move together with the velocity clearly lower than the velocity of light. Secondly, due to the efficient mixing,
the reconnection of the magnetic field should be very efficient. As a consequence, a relatively low value of the magnetic field is expected at the transition region. This magnetic field is described by the magnetization parameter of plasma $\sigma$ which in our case can have values clearly below that estimated for the Crab Nebula 
(i.e. $\sigma\sim 0.003$, Kennel \& Coroniti,~1984). Since the mixed winds move relatively slow, relativistic electrons
stay relatively long time in the dense radiation of the companion star losing energy on the IC scattering process. Thanks to the relatively weak magnetic field at the transition region, the cooling process of relativistic electrons is not completely dominated by synchrotron radiation as considered by e.g. Arons \& Tavani~(1993). 

As an example, we apply our model to the four MSP binary systems, two of the Black Widow type and two of the Redback type. We show that electrons can be accelerated in these binary systems to TeV energies. Significant amount of their energy is lost on the production of TeV $\gamma$-rays in collisions with the stellar radiation. We calculate the synchrotron and IC $\gamma$-ray spectra escaping to the observer located at different angles in respect to direction defined by the stars. Note that synchrotron spectra are independent of the location of the observer due to isotropization of relativistic electrons in the mixed winds and relatively low velocity of this winds. However, the IC $\gamma$-ray spectra clearly depend on the location of the observer since the companion star creates anisotropic radiation field for relativistic electrons. We have made simplified assumption that all relativistic electrons are injected at the apex of the collision region between the pulsar and stellar winds whose distance from the companion star is reported in Table.~1. In fact, the injection distance should extend along the transition region as discussed in Section.~2.  Therefore, in a more realistic injection scenario for electrons, the $\gamma$-ray spectra at the observer can be obtained by integrating over the specific solid angle subtracted by the extend of the transition region. However, such process will require introduction of a few, not well constrained, free parameters which should describe the extend of the injection region of electrons.
We do not consider these geometric effects in detail in the present calculations which are more oriented on the investigation of general emission features.

From normalization of calculated synchrotron spectra to the observed non-thermal X-ray flux from these binary systems (or to the upper limit in the case of PSR J1816+4510), we predict the absolute level of the TeV $\gamma$-ray flux from these binaries (see Figs.~3 and~4).
It is concluded that the TeV emission is comparable to the sensitivities of the present Cherenkov telescopes. Therefore, extensive observations of some MSP binary systems by the HESS, MAGIC or VERITAS telescopes should give a chance for positive detection of the new class of the TeV $\gamma$-ray sources.

Note that the expected TeV $\gamma$-ray emission from MSP binary systems should have different emission features than those discovered in the recent years from the massive TeV $\gamma$-ray binaries such as 
LS~5039 or LS~I~303~+61. After all, the maximum (minimum) of the TeV emission is predicted to appear at the range of phases when the pulsar is behind (in front) of the companion star. Moreover, we predict the synchronization of the GeV and TeV $\gamma$-ray emission in the light curve of the MSP binary systems. Note however, that predicted modulated GeV flux can not be observed with the present even extensive monitoring with the Fermi-LAT telescope, possibly except the cases of some Redback type binaries which contain especially luminous companion stars as observed in the case of PSR J1816+4510. 
These two basic emission features clearly distinguish MSP binary systems from the massive TeV binary systems. They are due to the much weaker radiation field created by the companion stars in the MSP binary systems when compared to the massive binaries. Therefore, also the IC $e^\pm$ pair cascading processes within the MSP binary systems are not expected to play a major role in contrast with the massive TeV binary systems. 

The results of our calculations base on the assumption that the winds from the pulsar and the companion star are spherically symmetric. In fact, this is likely not the case. The wind of the companion star is
probably much stronger towards the pulsar due to irradiation. On the other hand, also the pulsar wind is expected to be stronger in the equatorial plane in respect to the rotational axis of the pulsar
(see e.g. Bogovalov \& Khangoulian~2002, Lyubarsky~2002). Due to the non-spherical pulsar wind, the wind power can be more efficiently transferred to relativistic electrons and the predicted above $\gamma$-ray fluxes can be enhanced as considered also by Wu et al.~(2012).

\begin{acknowledgements}
This work is supported by the grants through the Polish NCN No. 2011/01/B/ST9/00411 and NCBiR No. ERA-NET-ASPERA/01/10.
\end{acknowledgements}

\end{document}